\documentclass{ws-ijmpa}
\usepackage[super,compress]{cite}
\usepackage{psfrag}
\usepackage{pstricks}

\def\vb#1{\vbox to #1 pt{}}
\def\slash#1{#1\!\!\! /}
\def\h#1{\hskip #1 mm}

\begin{document}


%
\catchline{}{}{}{}{}
%

\title{A resource for signs and Feynman diagrams of the Standard Model}

\author{Jorge~C.~Rom\~ao}

\address{Departamento de F\'\i sica and CFTP, Instituto Superior T\'ecnico\\
          Technical University of Lisbon, 1049-001 Lisboa, Portugal\\
jorge.romao@ist.utl.pt}

\author{Jo\~{a}o~P.~Silva}

\address{Instituto Superior de Engenharia de Lisboa,
	1959-007 Lisboa, Portugal\\
and\\
Centro de F\'{\i}sica Te\'{o}rica de Part\'{\i}culas (CFTP),
    Instituto Superior T\'{e}cnico, Technical University of Lisbon,
    1049-001 Lisboa, Portugal\\
jpsilva@cftp.ist.utl.pt}

\maketitle


\begin{abstract}
When performing a full calculation within the
Standard Model or its extensions,
it is crucial that one utilizes a consistent
set of signs for the gauge couplings and gauge fields.
Unfortunately, the literature is plagued with differing
signs and notations.
We present all Standard Model Feynman rules,
including ghosts,
in a convention-independent notation,
and we table the conventions in close to 40
books and reviews.

\end{abstract}

\ccode{PACS numbers:11.15.Bt, 12.10.Dm, 12.15.Lk}

\section{Introduction}	

Almost every book and review on the Standard Model (SM) has its own
conventions for the signs that enter the definitions of the couplings
and fields.  Although the signs are irrelevant when a full calculation
is made with any given convention, the signs of the various Feynman
diagrams are usually different in different conventions.  Of course,
most articles sidestep writing all Feynman diagrams, with the
rationale that these are already contained in several books.
Typically, an article on a model of Physics beyond the SM shows only a
few Feynman rules, or not even that.  As a result, the remaining
Feynman rules needed for any given calculation must be derived from
first principles or found in books.  And this is where the problem
resides; which convention was used in the article?  How does it
compare with the convention in some specific book?

Here we perform two tasks. We list all Feynman rules with arbitrary
signs, allowing one to specify later for any given sign convention
being used,
and we list the sign conventions of close to 40 known books and
reviews.

Section~\ref{sec:SM} summarizes the SM Lagrangian
including the generic signs (represented by parameters $\eta=\pm 1$)
necessary to specify the different notations found in the literature.
These are listed in table form in section~\ref{sec:compare},
including only those references we consulted which:
(i) follow the metric $(+,-,-,-)$;
(ii) follow Bjorken and Drell's \cite{BD} convention for the propagator,
with the explicit $i$;
and (iii) are internally consistent
(\textit{i.e.}, we do not include references which make
one sign choice in one part of the Lagrangian and a different choice elsewhere).
Sections~\ref{sec:FR_QCD} and \ref{sec:FR_EW} contain all Feynman rules of the SM,
including would-be Goldstone bosons and ghosts in an arbitrary $R_\xi$ gauge,
in a convention-independent notation.
Consistency remarks due Gauge invariance and
invariance under BRST transformations are relegated
to the \ref{app:BRST}.

\section{\label{sec:SM}The Standard Model}

\subsection{Gauge group $SU(3)_c$}

Here the important conventions are for the field strengths and the
covariant derivatives. We have
\begin{equation}
  \label{eq:apD-49}
  G_{\mu\nu}^a = \partial_{\mu} G_\nu^a -\partial_{\nu} G_\mu^a  - \eta_s
 g_s f^{abc} G_\mu^b G_\nu^c \quad (a=1,\ldots, 8) ,
\end{equation}
where $f^{abc}$ are the group structure constants, satisfying
\begin{equation}
  \label{eq:apD-50}
  \left[ T^a , T^b \right]= i f^{abc} T^c,
\end{equation}
and $T^a$ are the generators of the group.
The parameter $\eta_s = \pm 1$,
reflects the two usual signs in the literature.
The covariant derivative of
a (quark) field $q$ in some representation $T^a$ of the gauge group
is given by
\begin{equation}
  \label{eq:apD-90}
  D_\mu q = \left( \partial_\mu + i\,\eta_s\, g_s\, G^a_\mu T^a \right) q .
\end{equation}
In QCD,
the quarks are in the fundamental representation and $T^a=\lambda^a/2$,
where $\lambda^a$ are the Gell-Mann matrices.
A gauge transformation is given by a matrix
\begin{equation}
  \label{eq:92}
  U = e^{i\, \eta_s\,g_s\, T^a \beta^a},
\end{equation}
and the fields transform as
\begin{align}
\label{eq:121}
  &q \rightarrow e^{i\, \eta_s\, g_s\, T^a \beta^a} q ,
&&\delta q =i\, \eta_s\,g_s\, T^a \beta^a q ,
\nonumber\\[+2mm]
&G_\mu^a T^a \rightarrow U  G_\mu^a T^a U^{-1} + \frac{i}{\eta_s g_s} \partial_\mu U
  U^{-1} ,
&&\delta G_\mu^a = - \partial_\mu \beta^a -\eta_s\, g_s\,
  f^{abc} \beta^b G_\mu^c ,
\end{align}
where the second column is for infinitesimal transformations. With
these definitions one can verify that the covariant derivative
transforms like the field itself,
\begin{equation}
  \label{eq:apD-93}
  \delta (D_\mu q) = i\, \eta_s\,g_s\, T^a \beta^a (D_\mu q) ,
\end{equation}
ensuring the gauge invariance of the Lagrangian.
Further consistency checks due to gauge invariance will
be relegated to \ref{app:BRST}.

\subsection{Gauge Group $SU(2)_L \times U(1)_Y$}

For the $SU(2)_L$ group, we have
\begin{equation}
  \label{eq:apD-94}
  W_{\mu\nu}^a = \partial_{\mu} W_\nu^a -\partial_{\nu} W_\mu^a
  -\eta\, g  \epsilon^{abc} W_\mu^b W_\nu^c \quad (a=1,\ldots, 3) ,
\end{equation}
where, for the fundamental representation of $SU(2)_L$,
$T^a=\tau^a/2$, where $\tau^a$ are the Pauli matrices,
$\epsilon^{abc}$ is the completely anti-symmetric
tensor in 3 dimensions,
and $\eta=\pm 1$. 
The covariant derivative for any field $\psi_L$ transforming
non-trivially under this group is,
\begin{equation}
  \label{eq:ApD-99}
  D_\mu \psi_L = \left( \partial_\mu + i\,\eta\, g\, W^a_\mu T^a \right) \psi_L.
\end{equation}

As for the Abelian $U(1)_Y$ group,
we have
\begin{equation}
  \label{eq:apD-100}
  B_{\mu\nu} = \partial_{\mu} B_\nu -\partial_{\nu} B_\mu,
\end{equation}
with the covariant derivative given by
\begin{equation}
  \label{eq:apD-101}
  D_\mu \psi = \left( \partial_\mu +i\,\eta^\prime\, g^\prime\, \eta_Y Y\, B_\mu  \right) \psi,
\end{equation}
where $Y$ is the hypercharge of the field,
connected to the electric charge through
\begin{equation}
\label{eq:apD-102}
  Q = T_3 + \eta_YY\ .
\end{equation}
As before $\eta^\prime, \eta_Y = \pm 1$.
Some authors use
\begin{equation}
\label{eq:apD-102-theirs}
  Q = T_3 + \eta_Y \frac{Y_\textrm{theirs}}{2} = \frac{\tau_3 +\eta_Y
    Y_\textrm{theirs}}{2}, 
\end{equation}
instead of our Eq.~\eqref{eq:apD-102}.
The difference is immaterial for the Feynman rules,
which depend only on $Q$.

It is useful to write the covariant derivative in terms of the mass
eigenstates $A_\mu$ and $Z_\mu$.
These are defined by the relations\footnote{One
could also include a sign in the photon field $A$,
by substituting $A_\mu \rightarrow \eta_A A_\mu$.
However we have found no author who made the choice $\eta_A=-1$.}, 
\begin{equation}
  \label{eq:103}
  \left\{
    \begin{array}{l}
      W_\mu^3 = \eta_Z\, Z_\mu \cos\theta_W 
      + A_\mu \eta_\theta \sin\theta_W \\[+2mm]
      B_\mu= -\eta_Z Z_\mu \eta_\theta\,\sin\theta_W + A_\mu \cos\theta_W \\
    \end{array}
\right. , \quad
  \left\{
    \begin{array}{l}
    \eta_Z  Z_\mu = W_\mu^3 \cos\theta_W 
    - B_\mu \eta_\theta\,\sin\theta_W \\[+2mm]
      A_\mu=  W_\mu^3\, \eta_\theta \sin\theta_W + B_\mu \cos\theta_W \\
    \end{array}
\right. \ .
\end{equation}
For a doublet field $\psi_L$, with hypercharge $Y$, we get,
\begin{eqnarray}
D_\mu \psi_L
&=&
\left[ \partial_\mu + i\eta\, \frac{g}{\sqrt{2}} \left(
\tau^+ W_\mu^+ +  \tau^- W_\mu^- \right) +i\eta\, \frac{g}{2} \tau_3 W_\mu^3
+ i\, \eta^\prime g^\prime \eta_Y Y B_\mu \right] \psi_L
\nonumber\\[+2mm]
&=&
\left[ \partial_\mu + i\, \eta \frac{g}{\sqrt{2}} \left(
 \tau^+ W_\mu^+  +  \tau^- W_\mu^- \right) +i\, \eta_e e\, Q\,   A_\mu
\right.
\nonumber\\[+2mm]
& &
\left.
\ +\  i\, \eta \frac{g}{\cos\theta_W} \left(\frac{\tau_3}{2} - Q\, \sin^2\theta_W
\right) \eta_Z Z_\mu \right] \psi_L,
\label{eq:115}
\end{eqnarray}
where
\begin{eqnarray}
W_\mu^\pm &=&
\frac{W_\mu^1 \mp i W_\mu^2}{\sqrt{2}},
\label{eq:99}
\\
\tau_\pm &=&
\frac{\tau_1 \pm i \tau_2}{\sqrt{2}}.
\end{eqnarray}
The charge operator
is defined by
\begin{equation}
  \label{eq:104}
  Q = \left[
    \begin{matrix}
      \frac{1}{2} +\eta_Y Y & 0 \\[+2mm] 
      0&  -\frac{1}{2} +\eta_Y Y  \\[+2mm] 
    \end{matrix}
\right]\ ,
\end{equation}
and we have used the relations,
\begin{eqnarray}
  \eta_e \,e
&=&
(\eta\, \eta_\theta)\, g \sin\theta_W
\nonumber\\
&=& 
\eta^\prime\, g^\prime \cos\theta_W\, .
\label{eq:49}
\end{eqnarray}
Many authors use $\eta_e=+1$.
Some authors use $\eta_e=-1$,
to account for their other conventions
(notably $\eta=\eta^\prime=-1$),
and still keep $e =  + g^\prime \cos{\theta_W} = + g \sin{\theta_W}$.
For a singlet of $SU(2)_L$, $\psi_R$, we have,
\begin{eqnarray}
  D_\mu \psi_R
&=& \left[ \partial_\mu  
+ i\, \eta^\prime g^\prime\eta_Y Y B_\mu \right] \psi_R 
\nonumber\\[+2mm]
&=& \left[ \partial_\mu  +i\, \eta_e  e\, Q\,   A_\mu
- i\, \eta \frac{ g}{\cos\theta_W}  Q\, \sin^2\theta_W 
\eta_Z Z_\mu \right] \psi_R\, .
\label{eq:105}
\end{eqnarray}
We collect in Table~\ref{tab:SM} the quantum numbers of the SM
particles. 

\begin{table}[ph]
\tbl{Values of $T_3^f$, $Q$ and $Y$ for the SM particles.}
{\begin{tabular}{@{}cccccccccc@{}} \toprule
Field& $\ell_L$&$\ell_R$& $\nu_L$&  
$u_L$ &$d_L$ &$u_R$ &$d_R$ & $\phi^+$& $\phi^0$ 
\\[+2mm]
\colrule
$T_3$ & $-\tfrac{1}{2}$&$0$& $\tfrac{1}{2}$&$\tfrac{1}{2}$&
$-\tfrac{1}{2}$& $0$& $0$ &$\tfrac{1}{2}$&$-\tfrac{1}{2}$
\\[+2mm]
$\eta_Y Y$ &$-\tfrac{1}{2}$&$-1$&$-\tfrac{1}{2}$&$\tfrac{1}{6}$&$\tfrac{1}{6}$
&$\tfrac{2}{3}$&$-\tfrac{1}{3}$&$\tfrac{1}{2}$ &$\tfrac{1}{2}$
\\[+2mm]
$Q$&$-1$&$-1$&$0$&$\tfrac{2}{3}$&$-\tfrac{1}{3}$&$\tfrac{2}{3}$&$-\tfrac{1}{3}$
&$1$&$0$
\\[+2mm]
\botrule
\end{tabular} \label{tab:SM}}
\end{table}

Notice that the right-hand sides of
Eqs.~\eqref{eq:115} and \eqref{eq:105} only involve
$Y$ through $Q$,
where it appears in the combination $\eta_Y Y$.
A few authors write Eqs.~\eqref{eq:115} and \eqref{eq:105} directly for each field,
sidestepping a precise definition for their $\eta_Y$.

For each fermion field $\psi$,
one defines
$\psi_{R,L} = P_{R,L} \psi$,
where
\begin{equation}
P_{R,L} = \frac{1 \pm \gamma_5}{2},
\end{equation}
and $\psi = \psi_R + \psi_L$.

\subsection{The gauge and fermion fields Lagrangian}

The gauge field Lagrangian is
\begin{equation}
  \label{eq:113}
  \mathcal{L}_{\rm gauge} = -\frac{1}{4} G_{\mu\nu}^a G^{a \mu\nu}
-\frac{1}{4} W_{\mu\nu}^a W^{a \mu\nu}
-\frac{1}{4} B_{\mu\nu} B^{\mu\nu},
\end{equation}
where the field strengths are given in
Eqs.~(\ref{eq:apD-49}), (\ref{eq:apD-94}) and (\ref{eq:apD-100}).

The kinetic terms for the fermions,
including the interaction with the gauge fields due to the covariant derivative,
is written as
\begin{equation}
  \label{eq:114}
  \mathcal{L}_{\rm Fermion} =  
\sum_{\rm quarks} i \overline{q} \gamma^\mu D_\mu q 
+ \sum_{\psi_L} i \overline{\psi_L} \gamma^\mu  D_\mu \psi_L
+ \sum_{\psi_R} i \overline{\psi_R} \gamma^\mu  D_\mu \psi_R,
\end{equation}
where the covariant derivatives are obtained with the rules in
Eqs.~(\ref{eq:apD-90}),~(\ref{eq:115}) and (\ref{eq:105}).

\subsection{The Higgs Lagrangian}

The SM includes a Higgs doublet with the following assignments,
\begin{equation}
  \label{eq:50}
  \Phi = \left[
    \begin{matrix}
      \varphi^+ \\[+2mm]
      \displaystyle
       \frac{v + H + i \varphi_Z}{\sqrt{2}} \\
    \end{matrix}
\right].
\end{equation}
Since $\eta_Y Y_\Phi = + 1/2$,
the covariant derivative reads
\begin{eqnarray}
  D_\mu \Phi
&=& \left[ \partial_\mu + i\, \eta \frac{ g}{\sqrt{2}} \left(
 \tau^+ W_\mu^+ +  \tau^- W_\mu^- \right) +i\, \eta \frac{g}{2} \tau_3 W_\mu^3
+ i\, \eta' \frac{g'}{2} B_\mu \right] \Phi
\nonumber\\[+2mm]
&=&
\left[ \partial_\mu + i\,\eta \frac{g}{\sqrt{2}} \left(
\tau^+ W_\mu^+ +  \tau^- W_\mu^- \right) +i\, \eta_e e\, Q\,   A_\mu
\right.
\nonumber\\[+2mm]
& &
\left.
\ +\  i\,\eta \frac{g}{\cos\theta_W} \left(\frac{\tau_3}{2} - Q\, \sin^2\theta_W
\right)\eta_Z Z_\mu \right] \Phi,
\label{eq:90}
\end{eqnarray}
where, for the doublet field $\Phi$,
\begin{equation}
Q =
\left(
\begin{array}{cc}
1 & 0\\
0 & 0
\end{array}
\right).
\end{equation}
The Higgs Lagrangian is
\begin{equation}
  \label{eq:93}
\mathcal{L}_{\rm Higgs} =  \left(D_\mu \Phi\right)^\dagger D_\mu \Phi
+ \mu^2 \Phi^\dagger \Phi - \lambda \left(\Phi^\dagger \Phi\right)^2 ,
\end{equation}
leading to the relations,
\begin{equation}
  \label{eq:4}
  v^2 = \frac{\mu^2}{\lambda}, \quad m_h^2 = 2 \mu^2, \quad
 \lambda= \frac{g^2}{8} \frac{m_h^2}{m_W^2}.
\end{equation}
Expanding this Lagrangian,
we find the following terms quadratic in the fields:
\begin{eqnarray}
  \mathcal{L}_{\rm Higgs}
&=&
\cdots + 
\frac{1}{8} g^2 v^2 W_\mu^3 W^{\mu 3} + \frac{1}{8} g'^2 v^2 B_\mu B^{\mu}
- \frac{1}{4}\eta \eta' g g' v^2 W_\mu^3 B^{\mu} 
+ \frac{1}{4} g^2 v^2 W_\mu^+  W^{-\mu}
\nonumber\\[+2mm]
& &
\hspace{5mm} 
+ \frac{1}{2} v\, \partial^\mu \varphi_Z \left(\eta' g' B_\mu -\eta\, g W_\mu^{3}
\right) - \frac{i}{2}\eta\, g v W_\mu^- \partial^\mu \varphi^+ 
+ \frac{i}{2}\eta\, g v W_\mu^+ \partial^\mu \varphi^- \ .
\label{eq:94}
\end{eqnarray}
The first three terms give, after diagonalization, a massless field
(the photon), and a massive one (the $Z$), with the relations given in
Eq.~(\ref{eq:103}), while the fourth term gives mass to the charged
$W^\pm_\mu$ bosons.
Using Eq.~(\ref{eq:103}), we get
\begin{eqnarray}
\mathcal{L}_{\rm Higgs}
&=&
\cdots + 
\frac{1}{2} m_Z^2 Z_\mu Z^\mu + m_W^2 W_\mu^+  W^{-\mu} 
\nonumber\\[+2mm]
& &
\hspace{5mm}
- \eta\,\eta_Z\, m_Z Z_\mu \partial^\mu \varphi_Z 
- i\, \eta\, m_W \left( W_\mu^- \partial^\mu \varphi^+ -
W_\mu^+ \partial^\mu \varphi^-\right),
\label{eq:100}
\end{eqnarray}
where
\begin{equation}
  \label{eq:101}
  m_W = \frac{1}{2} g v, \quad
m_Z = \frac{1}{\cos\theta_W} \frac{1}{2} g v =
\frac{1}{\cos\theta_W} m_W.
\end{equation}
By looking at Eq.~(\ref{eq:100}) we realize that,
besides finding a realistic spectra for the gauge bosons,
we also get a problem.
In fact,
the terms in the last line are quadratic in the fields and complicate
the definition of the propagators.
The gauge fixing terms discussed in section~\ref{subsec:GFix}
solve this problem.

\subsection{The Yukawa Lagrangian, fermion masses and the CKM matrix}

After spontaneous symmetry breaking,
the interaction between the fermions and the Higgs doublet
gives masses to the elementary fermions.
We have,
\begin{equation}
\label{eq:116}
\mathcal{L}_{\rm Yukawa}=
- \overline{L}_L\,  Y_l\, \Phi\ \ell_R
- \overline{Q}^\prime_L\,  Y_d\,  \Phi\ d^\prime_R
- \overline{Q}^\prime_L\,  Y_u\,  \widetilde{\Phi}\ u^\prime_R +
\textrm{h.c.} ,
\end{equation}
where a sum over generations is implied by the matrix notation,
$L_L$ ($Q^\prime_L$) are the left-handed lepton (quark) doublets and,
\begin{equation}
  \label{eq:118}
  \widetilde{\Phi} = i\, \sigma_2 \Phi^* = \left[
    \begin{matrix}
\displaystyle
      \frac{v + H - i \varphi_Z}{\sqrt{2}}\\[+2mm]
      -\varphi^-
    \end{matrix}
\right].
\end{equation}
$Y_l$, $Y_d$, and $Y_u$ are general complex $3 \times 3$ matrices in
the respective flavor spaces.

To bring the quarks into the mass basis,
$Y_d$ and $Y_u$ are diagonalized through unitary transformations
\begin{eqnarray}
\overline{u}^\prime_L
= 
\overline{u}_L\, U^\dagger_{uL},
& \hspace{10mm} &
\overline{d}^\prime_L
=
\overline{d}_L\, U^\dagger_{dL},
\nonumber\\
u^\prime_R
=
U_{uR}\, u_R,
& \hspace{10mm} &
d^\prime_R
=
U_{dR}\, d_R,
\label{to_mass_basis}
\end{eqnarray}
such that
\begin{eqnarray}
\frac{v}{\sqrt{2}}\, U^\dagger_{uL}\, Y_u\, U_{uR}
&=&
M_u =
\textrm{diag} \left( m_u, m_c, m_t \right),
\nonumber\\
\frac{v}{\sqrt{2}}\, U^\dagger_{dL}\, Y_d\, U_{dR}
&=&
M_d =
\textrm{diag} \left( m_d, m_s, m_b \right).
\end{eqnarray}
In this new basis,
the Higgs couplings of the quarks become diagonal:
\begin{equation}
- \mathcal{L}_{\rm H}
=
\left(1 + \frac{h^0}{v} \right)
\left[
\overline{u}\,  M_u\, u
+
\overline{d}\,  M_d\, d
\right].
\end{equation}
The couplings to the photon and the $Z$ remain diagonal.
In contrast,
the couplings to the $W$ mix the upper and lower components
of $Q^\prime_L$,
which transform differently under Eqs.~\eqref{to_mass_basis}.
As a result,
the couplings to $W^\pm$ become off-diagonal:
\begin{equation}
- \eta\, \mathcal{L}_{\rm W}
=
\frac{g}{\sqrt{2}}\,
\overline{u}_L\, V\, \gamma^\mu\, d_L\, W_\mu^\dagger 
+
\textrm{h.c.} ,
\end{equation}
where
\begin{equation}
V = U^\dagger_{uL} U_{dL}
\end{equation}
is the Cabibbo-Kobayashi-Maskawa (CKM) matrix,
which also affects the interactions with the charged
Goldstone bosons.

In the SM,
there are no right-handed neutrinos.
As a result,
the neutrinos are massless and we are free to
rotate them in order to accommodate the
transformations of the charged quarks needed
to diagonalize $Y_l$.
Thus,
without loss of generality,
we may take $Y_l = \textrm{diag} \left( m_e, m_\mu, m_\tau \right)$
and $V=1$ in the leptonic sector.

\subsection{\label{subsec:GFix}The gauge fixing}

One needs to gauge fix the gauge part of the
Lagrangian in order to be able to define the propagators.
In the $R_\xi$ gauges, the gauge fixing Lagrangian reads:
\begin{equation}
  \label{eq:102}
  \mathcal{L}_{\rm GF} = -\frac{1}{2\xi_G} F_G^2 -\frac{1}{2\xi_A} F_A^2 
-\frac{1}{2\xi_Z} F_Z^2 - \frac{1}{\xi_W} F_{-} F_{+} ,
\end{equation}
where
\begin{eqnarray}
F_G^{a}
&=& \partial^\mu  G_\mu^{a},
\nonumber\\
F_A
&=&
\partial^\mu A_\mu ,
\nonumber\\
F_Z 
&=&
\partial^\mu Z_\mu +\eta\,\eta_Z\, \xi_Z m_Z \varphi_Z,
\nonumber\\
F_{+} &=&
\partial^\mu W_\mu^{+} +i\,\eta\, \xi_W m_W \varphi^{+} ,
\nonumber\\
F_{-} &=&
\partial^\mu W_\mu^{-} - i\,\eta\, \xi_W m_W \varphi^{-}.
\label{eq:106}
\end{eqnarray}
One can easily verify that,
with these definitions,
$\mathcal{L}_{\rm GF}$ cancels the mixed quadratic terms
on the second line of Eq.~(\ref{eq:100}).

\subsection{The ghost Lagrangian}

The last piece needed for the SM Lagrangian is the ghost Lagrangian.
For a linear gauge fixing condition, as in Eq.~(\ref{eq:106}), this is
given by the Fadeev-Popov prescription:
\begin{eqnarray}
\mathcal{L}_{\rm Ghost}
&=&
\eta_G\,
\sum_{i=1}^4 \left[ 
\overline{c}_{+} \frac{\partial(\delta F_{+})}{\partial \alpha^i}
+\overline{c}_{-} \frac{\partial(\delta F_{+})}{\partial \alpha^i}
+\overline{c}_{Z} \frac{\partial(\delta F_{Z})}{\partial \alpha^i}
+\overline{c}_{A} \frac{\partial(\delta F_{A})}{\partial \alpha^i}
\right] c_i
\nonumber\\[+2mm]
& &
+\ 
\eta_G\,
\sum_{a,b=1}^8 \overline{\omega}^a \frac{\partial(\delta
F_{G}^a)}{\partial \beta^b} \omega^b,
\label{eq:107} 
\end{eqnarray}
where we have denoted by $\omega^a$ the ghosts associated with the
$SU(3)_c$ transformations defined by Eq.~\eqref{eq:92},
and by $c_\pm,c_A,c_Z$ the electroweak ghosts associated with the
gauge transformations,
\begin{equation}
\label{eq:109a}
U = e^{i\, \eta\, g T^a \alpha^a} \quad (a=1,\ldots,3),
\end{equation}
and
\begin{equation}
\label{eq:109b}
U = e^{i\,\eta'\eta_Y\, g' Y \alpha^4}.
\end{equation}
For completeness,
we write in \ref{app:BRST} the gauge transformations of the gauge
fixing terms needed to find the Lagrangian in Eq.~(\ref{eq:107}).

Because ghosts are not external states,
the sign $\eta_G = \pm 1$ is immaterial and,
although it corresponds to an overall sign
affecting all propagators and vertices with ghosts,
it drops out in any physical calculation involving ghosts.

\subsection{The complete SM Lagrangian}

Finally,
the complete Lagrangian for the Standard Model is obtained
putting together all the pieces. We have,
\begin{equation}
  \label{eq:117}
  \mathcal{L}_{\rm SM}= \mathcal{L}_{\rm gauge} +\mathcal{L}_{\rm
    Fermion} + \mathcal{L}_{\rm Higgs} + \mathcal{L}_{\rm Yukawa} + 
\mathcal{L}_{\rm GF}+ \mathcal{L}_{\rm Ghost} ,
\end{equation}
where the different terms were given in
Eqs.~(\ref{eq:113}),~(\ref{eq:114}),~(\ref{eq:93}),~(\ref{eq:116}),~(\ref{eq:102}),
and (\ref{eq:107}).

\section{\label{sec:compare}Notations found in the literature}

In order to use the results contained in some specific source in the literature,
one must find the covariant derivative
\begin{equation}
D_\mu =
\partial_\mu
+ i \eta\, g \frac{\tau_a}{2} W_\mu^a
+ i \eta^\prime\, \eta_Y\, g^\prime Y\, B_\mu,
\label{Dmu}
\end{equation}
and Eqs.~\eqref{eq:apD-102} and \eqref{eq:103}.
This sets the sign convention for $\eta$,
$\eta^\prime$,
$\eta_Z$,
$\eta_\theta$,
and $\eta_Y$.
Typically, authors set $\eta_s=\eta$.

The signs and conventions in the literature are shown in Table~\ref{tab:signs}.

\begin{table}[ht]
\tbl{Sign conventions found in the literature. An asterisk, $\ast$, on the
last column means that such authors have $Q = (\tau_3 + Y_{\rm theirs})/2$
instead of our Eq.~\eqref{eq:apD-102}.}
{\begin{tabular}{@{}lccccccc@{}} \toprule
Ref. 
&\h{5} $\eta$ \h{5} &\h{5}  $\eta^\prime$\h{5} &\h{5} $\eta_Z$\h{5} &\h{5} $\eta_\theta$ \h{5}& \h{5}$\eta_Y$\h{5} & \h{5}$\eta_e$ \h{5}& $Y$\\
\colrule
\citen{Bailin, Pok, Prague, Pal, MS, Paul} &
\h{5} $+$ \h{5} & \h{5} $+$ \h{5} & \h{5} $+$ \h{5} & \h{5} $+$ \h{5} & \h{5} $+$ \h{5} & \h{5} $+$ \h{5} & 
\\
\citen{Quigg, HM, Rivers, AH, HHG, Barger, Huang, DGH, Sterman, FGross, Gri} &
\h{5} $+$ \h{5} & \h{5} $+$ \h{5} & \h{5} $+$ \h{5} & \h{5} $+$ \h{5} & \h{5} $+$ \h{5} & \h{5} $+$ \h{5} & $\ast$
\\
\citen{PS, Gaume} & \h{5} $-$ \h{5} & \h{5} $-$ \h{5} & \h{5} $+$ \h{5} & \h{5} $+$ \h{5} & \h{5} $+$ \h{5} & \h{5} $-$ \h{5} & 
\\
\citen{Aoki, Okun, Ryder, CL, Mohapatra, Fayya, Rolnick, Kane, Nair, Djouadi, Zee} &
\h{5} $-$ \h{5} & \h{5} $-$ \h{5} & \h{5} $+$ \h{5} & \h{5} $+$ \h{5} & \h{5} $+$ \h{5} & \h{5} $-$ \h{5} & $\ast$
\\
\citen{BLS, Grimus:2007if}  &
\h{5} $-$ \h{5} & \h{5} $-$ \h{5} & \h{5} $+$ \h{5} & \h{5} $-$ \h{5} & \h{5} $+$ \h{5} & \h{5} $+$ \h{5} & 
\\
\refcite{IZ} &
\h{5} $-$ \h{5} & \h{5} $-$ \h{5} & \h{5} $-$ \h{5} & \h{5} $+$ \h{5} & \h{5} $+$ \h{5} & \h{5} $-$ \h{5} & $\ast$
\\
\refcite{sakakibara} &
\h{5} $-$ \h{5} & \h{5} $+$ \h{5} & \h{5} $+$ \h{5} & \h{5} $-$ \h{5} &  & \h{5} $+$ \h{5} &
\\
\citen{hZg, paulo} &
\h{5} $-$ \h{5} & \h{5} $+$ \h{5} & \h{5} $+$ \h{5} & \h{5} $-$ \h{5} & \h{5} $-$ \h{5} & \h{5} $+$ \h{5} &
\\
\refcite{Romao}  &
\h{5} $-$ \h{5} & \h{5} $+$ \h{5} & \h{5} $+$ \h{5} & \h{5} $-$ \h{5} & \h{5} $+$ \h{5} & \h{5} $+$ \h{5} & 
\\
\refcite{Das}  &
\h{5} $+$ \h{5} & \h{5} $-$ \h{5} & \h{5} $+$ \h{5} & \h{5} $-$ \h{5} & \h{5} $+$ \h{5} & \h{5} $-$ \h{5} & $\ast$
\\
\botrule
\end{tabular} \label{tab:signs}}
\end{table}

The corresponding Feynman rules are presented in the following
sections.
A few remarks are in order.
As mentioned,
since the Feynman rules depend only on $Q$,
authors may choose to sidestep a definition of
$\eta_Y$;
or whether they are using $Y$,
from Eq.~\eqref{eq:apD-102},
or $Y_\textrm{theirs}$,
from Eq.~\eqref{eq:apD-102-theirs};
or even neglect to mention the hypercharge $Y$ altogether.
Similarly,
$\eta_\theta$,
$\eta^\prime$ and $g^\prime$ are absent from the Feynman rules and,
thus, not needed in any calculation.
We see that only $\eta_s$,
in the strong sector,
and $\eta$,
$\eta_e$,
and $\eta_Z$,
in the electroweak sector,
show up in Feynman diagrams.

The fact that authors differ by their $\eta$ sign,
but all keep to the definition of $m_W$ and $m_Z$
in Eq.~\eqref{eq:101} (assumed positive),
means that diagrams proportional to gauge boson masses
are also affected by the sign choice.
Some conventions lead to peculiar results.
For example,
the convention in Ref.~\refcite{BLS} leads to
the unconventional $e = -g^\prime \cos{\theta_W}$,
while keeping the usual $e = g \sin{\theta_W}$.
If one wishes to keep all quantities positive in the
relation $m_Z=g v/(2 \cos{\theta_W})$,
then one must assume that $g^\prime$ is negative.
This is irrelevant for the Feynman rules,
where $g^\prime$ does not show,
but unusual.

The relevant electroweak choices for $\eta$, $\eta_e$, and $\eta_Z$
may be inferred from any given reference, as long as a few Feynman
rules are given.  For example, the coupling of the photon with
fermions (or $W^+ W^-$, or $\varphi^+ \varphi^-$) sets $\eta_e$.
Similarly, the coupling of the $Z$ with fermions (or $W^+ W^-$, or
$\varphi^+ \varphi^-$) sets $\eta \eta_Z$.  Finally, the coupling of
the $W^+$ with fermions sets $\eta$.
This sets the notation for all other Feynman rules, even when
Goldstone bosons and/or ghosts are included, except for $\eta_G$ which
can be found in any of the propagators or vertices involving ghosts.
The sign for $\eta_G$ is shown in Table~\ref{tab:eta_G} for those
references including ghosts.  A star (*) indicates the references
that only include Feynman rules with ghosts for the pure non-abelian
gauge theory or that have an incomplete list of the Feynman rules for
the electroweak ghosts. A dagger ($\dagger$) indicates the references that
include all Feynman rules, including electroweak ghosts.

\begin{table}[ht]
\tbl{Sign convention for $\eta_G$ found in the literature.}
{\begin{tabular}{@{}lc@{}} \toprule
Ref.
&\h{10}  $\eta_G$\h{10} \\
\colrule
\citen{MS, Rivers,
HHG, DGH, Sterman, FGross, PS, Ryder, Nair, Zee, 
IZ, sakakibara, hZg, Das}
&
\h{10} 
$+$$\h{3}\ast$
\h{10}
\\
\citen{Bailin, Aoki, BLS, paulo, Romao}
&
\h{10} 
$+$$\h{3}\dagger$ 
\h{10}
\\
\citen{Huang, CL, Paul}
&
\h{10}
$-$$\h{3}\ast$
\h{10}
\\
\citen{Pok}
&
\h{10}
$-$$\h{3}\dagger$ 
\h{10}
\\
\botrule
\end{tabular} \label{tab:eta_G}}
\end{table}

Next we present all Feynman rules, including the generic signs, which
have been obtained using the \texttt{FeynRules}
package~\cite{FeynRules}.  The first Roman letters $(a,d,c,d,e)$
denote group indices; the Roman letters $(i,j)$ refer to the QCD
component; the Greek letters $(\mu, \nu, \sigma, \rho)$ denote Lorentz
indices; while the first Greek letters $(\alpha, \beta)$, appearing in
the CKM matrix $V$, refer to the flavor indices.

We finish this section by comparing our results with those found in
the literature. We only do this comparison for the set of references that have
all the Feynman rules for the Standard Model, including ghosts,
namely, Refs.~\citen{Bailin, Pok, Aoki, BLS, paulo} and \refcite{Romao}.
We agree with Ref.~\refcite{Bailin} (including the errata) except for an
overall sign in Eqs.~(14.66) and (14.67).  As for Ref.~\refcite{Pok}, we
disagree with the four gluon vertex on page~572, but we agree when it
is written on page~557.  We 
also note that this reference has the complete Feynman rules
for the counterterms that we do not include here. Ref.~\refcite{Aoki}
has all the Feynman rules, including also those for the counterterms. The
conventions of this reference are different from all those that we
cite and therefore difficult to compare. However, we have checked a
reasonable number of Feynman rules and got agreement in all cases.
Ref.~\refcite{paulo} has all Feynman rules correct,
except for an overall sign on the last vertex on page A.16 and the fourth
on page A.18. We agree with all Feynman rules contained in
Refs.~\refcite{BLS} and \refcite{Romao}.

\section{\label{sec:FR_QCD}Feynman Rules for QCD}

We give separately the Feynman Rules for QCD and the electroweak part
of the Standard Model. All moments are incoming, except in the ghost
vertices where they are explicitly shown.

\subsection{Propagators}

\vspace{5mm}

\vbox{
\begin{equation}
\label{eq:119}
\hskip 5 true cm  -i \delta_{ab}\left[ \frac{g_{\mu \nu}}{k^2 + i \epsilon}
- (1-\xi_G) \frac{k_\mu k_\nu}{(k^2)^2}\right] 
\end{equation}
\begin{picture}(0,0)
\psfrag{m}{$\hskip -5mm \mu,a$}
\psfrag{n}{$\nu,b$}
\psfrag{G}{$g$}
\put(5,6){\includegraphics[width=3.5cm]{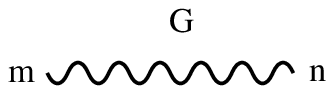}}
\end{picture}
}

\vspace{5mm}

\vbox{
\begin{equation}
\hskip 4 true cm  \delta_{ab}\ \frac{i\,\eta_G}{k^2 + i \epsilon}
\label{(5.7ab)}
\end{equation}
\begin{picture}(0,0)
\psfrag{G}{$\omega$}
\psfrag{a}{$a$}
\psfrag{b}{$b$}
\put(5,6){\includegraphics[width=3.5cm]{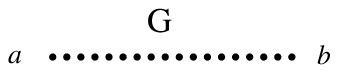}}
\end{picture}
}

\vspace{-3mm}

\subsection{Triple Gauge Interactions}

\vspace{8mm}

\vbox{
\begin{equation}
\label{eq:apD-GGG}
\hskip 50mm
\begin{array}{ll}
- \eta_s\, g_s f^{a b c} [ & g^{\mu \nu} (p_1 - p_2)^{\rho} + 
g^{\nu \rho} (p_2 - p_3)^{\mu}  \cr
\vb{20}&+ g^{\rho \mu} (p_3 - p_1)^{\nu} ]
\end{array}
\end{equation}
\psfrag{m,a}{$\mu,a$}
\psfrag{n,b}{$\nu,b$}
\psfrag{r,c}{$\rho,c$}
\psfrag{p1}{$p_1$}
\psfrag{p2}{$p_2$}
\psfrag{p3}{$p_3$}
\rput(2,1.2){\includegraphics{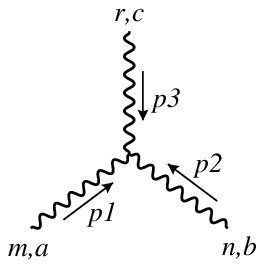}}
}

\subsection{Quartic Gauge Interactions}

\vspace{4mm}

\vbox{
\begin{equation}
\hskip 45mm
\begin{array}{ll}
- i g_s^2 \Big[ &f_{e a b} f_{e c d} (g_{\mu \rho} g_{\nu 
\sigma} - g_{\mu \sigma} g_{\nu \rho}) \cr
\vb{20}&+ f_{e a c} f_{e d b} (g_{\mu \sigma} g_{\rho \nu} - g_{\mu \nu} 
g_{\rho 
\sigma}) \cr
\vb{20}&+ f_{e a d} f_{e b c} (g_{\mu \nu} g_{\rho \sigma} - g_{\mu \rho} 
g_{\nu 
\sigma}) \Big]
\end{array}
\label{eq-c6.191}
\end{equation}
\psfrag{m}{$\mu,a$}
\psfrag{n}{$\nu,b$}
\psfrag{r}{$\rho,c$}
\psfrag{s}{$\sigma,d$}
\psfrag{p1}{$p_1$}
\psfrag{p2}{$p_2$}
\psfrag{p3}{$p_3$}
\psfrag{p4}{$p_4$}
\rput(2,1.8){\includegraphics{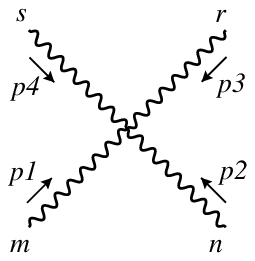}}
}

\vspace{-4mm}

\subsection{Fermion Gauge Interactions}

\vspace{13mm}

\vbox{
\begin{equation}
\hskip 50mm -i\,\eta_s\, g_s \gamma^{\mu} T^a_{i j}
\label{eq:apD-FFG} 
\end{equation}
\psfrag{m}{$\mu,a$}
\psfrag{a}{$j$}
\psfrag{b}{$i$}
\psfrag{p1}{$p_1$}
\psfrag{p2}{$p_2$}
\psfrag{p3}{$p_3$}
\rput(2,1){\includegraphics{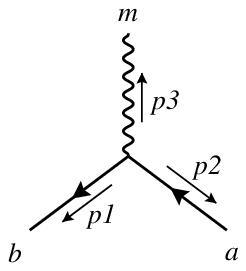}}
}

\subsection{Ghost Interactions}

\vspace{13mm}

\vbox{
\begin{equation}
\hskip 50mm
\begin{array}{ll}
&-\eta_s\eta_G\, g_s\ f^{a b c} p^{\mu}_1
\end{array}
\label{eq-c6.192}
\end{equation}
\psfrag{m}{$\mu,c$}
\psfrag{a}{$a$}
\psfrag{b}{$b$}
\psfrag{p1}{$p_1$}
\psfrag{p2}{$p_2$}
\psfrag{p3}{$p_3$}
\rput(2,1){\includegraphics{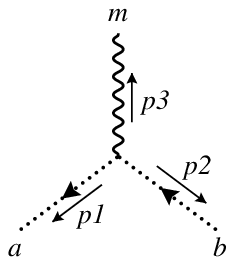}}
}

\section{\label{sec:FR_EW}Feynman Rules for the Electroweak Theory}

\subsection{Propagators}

\vspace{2mm}

\vbox{
\begin{equation}
\hskip 4 true cm   -i \left[ \frac{g_{\mu \nu}}{k^2 + i \epsilon}
- (1-\xi_A) \frac{k_\mu k_\nu}{(k^2)^2}\right] 
\end{equation}
\begin{picture}(0,0)
\psfrag{m}{$\mu$}
\psfrag{n}{$\nu$}
\psfrag{G}{$\gamma$}
\put(2,6){\includegraphics[width=3.5cm]{itcc5f1abc-new.eps}}
\end{picture}
}

\vspace{3mm}

\vbox{
\begin{equation}
  \hskip 42mm  -i\,\frac{1}{k^2 -m^2_W + i \epsilon} 
 \left[ g_{\mu \nu}- (1-\xi_W) \frac{k_\mu k_\nu}{k^2 -\xi_W m^2_W} \right] 
\label{(5.7bb)}
\end{equation}
\begin{picture}(0,0)
\psfrag{m}{$\mu$}
\psfrag{n}{$\nu$}
\psfrag{G}{$W$}
\put(2,6){\includegraphics[width=3.5cm]{itcc5f1abc-new.eps}}
\end{picture}
}

\vspace{3mm}

\vbox{
\begin{equation}
\hskip 44mm   -i\,\frac{1}{k^2 -m^2_Z + i \epsilon} 
 \left[ g_{\mu \nu}- (1-\xi_Z) \frac{k_\mu k_\nu}{k^2-\xi_Z m^2_Z} \right] 
\label{(5.7cb)}
\end{equation}
\begin{picture}(0,0)
\psfrag{m}{$\mu$}
\psfrag{n}{$\nu$}
\psfrag{G}{$Z$}
\put(2,6){\includegraphics[width=3.5cm]{itcc5f1abc-new.eps}}
\end{picture}
}

\vspace{3mm}

\vbox{
\begin{equation}
\hskip 35mm   \frac{i (\slash{p}+m_f)}{p^2-m_f^2+i \epsilon} 
\label{(5.12b)}
\end{equation}
\begin{picture}(0,0)
\psfrag{p}{$p$}
\put(2,2){\includegraphics[width=3.5cm]{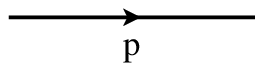}}
\end{picture}
}

\vspace{3mm}

\vbox{
\begin{equation}
\hskip 35mm   \frac{i}{p^2-m_h^2+i \epsilon} 
\end{equation}
\begin{picture}(0,0)
\psfrag{p}{$p$}
\psfrag{H}{$h$}
\put(2,-2.8){\includegraphics[width=3.5cm]{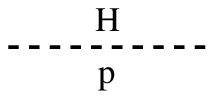}}
\end{picture}
}

\vspace{3mm}
\vbox{
\begin{equation}
\hskip 35mm   \frac{i}{p^2-\xi_Z m_Z^2+i \epsilon} 
\end{equation}
\begin{picture}(0,0)
\psfrag{p}{$p$}
\psfrag{H}{$\varphi_Z$}
\put(2,-2.5){\includegraphics[width=3.5cm]{HiggsPropagator.eps}}
\end{picture}
}

\vspace{3mm}

\vbox{
\begin{equation}
\hskip 35mm   \frac{i}{p^2- \xi_W m_W^2+i \epsilon} 
\end{equation}
\begin{picture}(0,0)
\psfrag{p}{$p$}
\psfrag{H}{$\varphi^\pm$}
\put(2,-2.5){\includegraphics[width=3.5cm]{HiggsPropagator.eps}}
\end{picture}
}

\subsection{Triple Gauge Interactions}

\vspace{15mm}

\vbox{
\begin{equation}
  \label{eq:190}
\hskip 32mm
   -i\, \eta_e\, e \left[ g_{\sigma \rho} (p_{-} - p_{+})_\mu
 +g_{\rho\mu} (p_{+}-q)_{\sigma} + g_{\mu\sigma} (q-p_{-})_{\rho}\right]
\end{equation}
\begin{picture}(0,18)
\psfrag{p}{$\h{-2.5} p_{-}$}
\psfrag{q}{$q$}
\psfrag{k}{$\h{-2.5} p_{+}$}
\psfrag{W1}{$W_{\sigma}^-$}
\psfrag{W2}{$W_{\rho}^+$}
\psfrag{A}{$A_{\mu}$}
\put(0.9,10){\includegraphics[width=30mm]{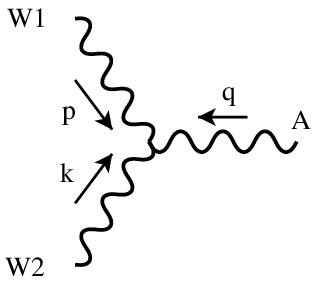}}
\end{picture}
}

\vspace{3mm}

\vbox{
\begin{equation}
  \label{eq:191}
\hskip 29.5mm
  - i\eta\eta_Zg\! \cos\theta_W\!\! \left[ 
    g_{\sigma\rho} (p_{-}\!\! -\! p_{+})_\mu
  \!+\!g_{\rho\mu} (p_{+}\!\!- \!q)_{\sigma} \!+\! g_{\mu\sigma}
 (q\!-\!p_{-})_{\rho}\right] 
\end{equation}
\begin{picture}(0,5)
\psfrag{p}{$\h{-2.5} p_{-}$}
\psfrag{q}{$q$}
\psfrag{k}{$\h{-2.5} p_{+}$}
\psfrag{W1}{$W_{\sigma}^-$}
\psfrag{W2}{$W_{\rho}^+$}
\psfrag{Z}{$\hskip -2mm Z_{\mu}$}
\put(0.9,-4.5){\includegraphics[width=30mm]{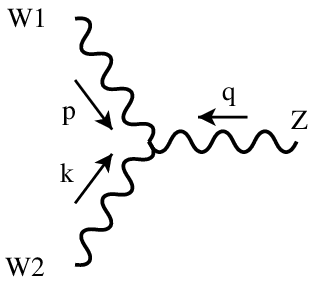}}  
\end{picture}
}

\subsection{Quartic Gauge Interactions}

\vspace{10mm}

\vbox{
\begin{equation}
  \label{eq:192}
\hskip 30mm
 - i e^2 \left[ 2 g_{\sigma\rho} g_{\mu\nu} -
g_{\sigma\mu} g_{\rho\nu} -g_{\sigma\nu} g_{\rho\mu} \right]
\end{equation}
\begin{picture}(0,5)
\psfrag{W1}{$\!\!W_{\sigma}^+$}
\psfrag{W2}{$\!\!A_{\mu}$}
\psfrag{W3}{$W_{\rho}^-$}
\psfrag{W4}{$A_{\nu}$}
\put(2,-5){\includegraphics[width=35mm]{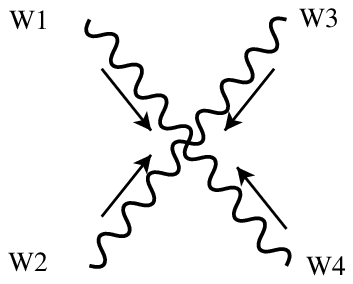}}
\end{picture}
}

\vspace{15mm}

\vbox{
\begin{equation}
\hskip 30mm
  -i g^2\, \cos^2\theta_W \left[ 2 g_{\sigma\rho} g_{\mu\nu} -
g_{\sigma\mu} g_{\rho\nu} -g_{\sigma\nu} g_{\rho\mu} \right]
\end{equation}
\begin{picture}(0,4)
\psfrag{W1}{$\!\!W_{\sigma}^+$}
\psfrag{W2}{$\!\!Z_{\mu}$}
\psfrag{W3}{$W_{\rho}^-$}
\psfrag{W4}{$Z_{\nu}$}
\put(2,-5){\includegraphics[width=35mm]{WPWMWPWM.eps}}
\end{picture}
}

\vspace{15mm}

\vbox{
\begin{equation}
\hskip 40mm
  -i\,\eta_e \eta\, \eta_Z\, e g \cos\theta_W \left[ 2 g_{\sigma\rho}
  g_{\mu\nu} - g_{\sigma\mu} g_{\rho\nu} -g_{\sigma\nu} g_{\rho\mu} \right]
\end{equation}
\begin{picture}(0,5)
\psfrag{W1}{$\!\!W_{\sigma}^+$}
\psfrag{W2}{$\!\!A_{\mu}$}
\psfrag{W3}{$W_{\rho}^-$}
\psfrag{W4}{$Z_{\nu}$}
\put(2,-5){\includegraphics[width=35mm]{WPWMWPWM.eps}}
\end{picture}
}

\vspace{15mm}

\vbox{
\begin{equation}
\hskip 30mm
  i g^2 \left[ 2 g_{\sigma\mu} g_{\rho\nu} -
g_{\sigma\rho} g_{\mu\nu} -g_{\sigma\nu} g_{\rho\mu} \right]
\end{equation}
\begin{picture}(0,5)
\psfrag{W1}{$\!\!W_{\sigma}^+$}
\psfrag{W3}{$\!\!W_{\rho}^-$}
\psfrag{W2}{$W_{\mu}^+$}
\psfrag{W4}{$W_{\nu}^-$}
\put(2,-5){\includegraphics[width=35mm]{WPWMWPWM.eps}}
  \end{picture}
}

\subsection{Charged Current Interaction}

\vspace{10mm}

\vbox{
\begin{equation}
  \label{eq:188a}
\hskip 20mm
\displaystyle -i\,\eta\,
\frac{g}{\sqrt{2}}\gamma_{\mu}P_L\, V_{\alpha \beta}
\end{equation}
\begin{picture}(0,5)
\psfrag{x1}{$d_\beta$}
\psfrag{x2}{$u_\alpha$}
\psfrag{W}{$W^+_{\mu}$}
\put(20,0){\includegraphics[width=28mm,clip]{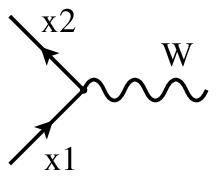}}
\end{picture}
}

\vspace{6mm}

\vbox{
\begin{equation}
  \label{eq:188b}
\hskip 20mm
\displaystyle -i\,\eta\,
\frac{g}{\sqrt{2}}\gamma_{\mu}P_L\, V^\ast_{\alpha \beta}
\end{equation}
\begin{picture}(0,5)
\psfrag{x1}{$u_\alpha$}
\psfrag{x2}{$d_\beta$}
\psfrag{W}{$W^-_{\mu}$}
\put(20,0){\includegraphics[width=28mm,clip]{CorrenteCarregada-signs.eps}}
\end{picture}
}

\vspace{6mm}

\vbox{
\begin{equation}
  \label{eq:188c}
\hskip 20mm
\displaystyle -i\,\eta\, 
\frac{g}{\sqrt{2}}\gamma_{\mu}P_L
\end{equation}
\begin{picture}(0,5)
\psfrag{x1}{$\ell, \nu$}
\psfrag{x2}{$\nu, \ell$}
\psfrag{W}{$W^\pm_{\mu}$}
\put(20,0){\includegraphics[width=28mm,clip]{CorrenteCarregada-signs.eps}}
\end{picture}
}

\subsection{Neutral Current Interaction}

\vspace{10mm}

\vbox{
\begin{equation}
\hskip 10mm
- i\,\eta_e\, e Q_f \gamma_{\mu}
\end{equation}
\begin{picture}(0,5)
\psfrag{x3}{$\psi_{f}$}
\psfrag{Z}{$Z_{\mu}$}
\psfrag{A}{$A_{\mu}$}
\put(20,-1){\includegraphics[width=27mm]{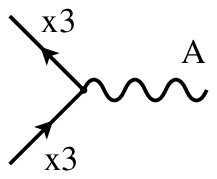}}
\end{picture}
}

\vspace{6mm}

\vbox{
\begin{equation}
\hskip 20mm
-i\,\eta\,\eta_Z\, 
\frac{g}{\cos\theta_W}\gamma_{\mu}\left(g_V^f-g_A^f \gamma_5\right)
\end{equation}
\begin{picture}(0,5)
\psfrag{x3}{$\psi_{f}$}
\psfrag{Z}{$Z_{\mu}$}
\psfrag{A}{$A_{\mu}$}
\put(20,0){\includegraphics[width=27mm]{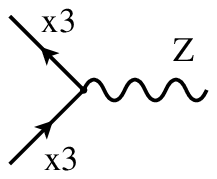}}
\end{picture}
}

\noindent
where
\begin{equation}
g_V^f = \frac{1}{2} T^3_f - Q_f \sin^2\theta_W, \quad
g_A^f = \frac{1}{2} T^3_f \ .
\end{equation}

\subsection{Fermion-Higgs and Fermion-Goldstone 
Interactions}

\vbox{
\vspace{12mm}
\begin{equation}
\hskip 10mm
  -i\, \displaystyle\frac{g}{2}\, \displaystyle
\frac{m_f}{m_W}
\end{equation}
\begin{picture}(0,5)
\psfrag{S}{$h$}
\psfrag{f1}{$f$}
\psfrag{f2}{$f$}
\put(20,-3){\includegraphics[width=30mm]{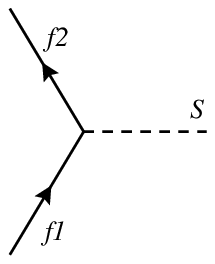}}  
\end{picture}
}

\vbox{
\vspace{5mm}
\begin{equation}
\hskip 10mm
   - g\, T^3_f\, \frac{m_f}{m_W}\, \gamma_5
\end{equation}
\begin{picture}(0,5)
\psfrag{S}{$\varphi_Z$}
\psfrag{f1}{$f$}
\psfrag{f2}{$f$}
\put(20,-4){\includegraphics[width=30mm]{PsiPsiHiggs.eps}  }
\end{picture}
}

\vbox{
\vspace{15mm}
\begin{equation}
  \label{eq:195a}
\hskip 30mm 
 i \displaystyle
  \frac{g}{\sqrt{2}}\left(\frac{m_{u \alpha}}{m_W} P_L -
    \frac{m_{d \beta}}{m_W}P_R \right)\, V_{\alpha \beta}
\end{equation}
\begin{picture}(0,5)
\psfrag{S}{$\varphi^+$}
\psfrag{f2}{$u_\alpha$}
\psfrag{f1}{$d_\beta$}
\put(20,-3){\includegraphics[width=30mm]{PsiPsiHiggs.eps}}
\end{picture}
}

\vbox{
\vspace{15mm}
\begin{equation}
  \label{eq:195b}
\hskip 30mm 
 i \displaystyle
  \frac{g}{\sqrt{2}}\left(\frac{m_{u \alpha}}{m_W}P_R -
    \frac{m_{d \beta}}{m_W}P_L \right)\, V^\ast_{\alpha \beta}
\end{equation}
\begin{picture}(0,5)
\psfrag{S}{$\varphi^-$}
\psfrag{f2}{$d_\beta$}
\psfrag{f1}{$u_\alpha$}
\put(20,-3){\includegraphics[width=30mm]{PsiPsiHiggs.eps}}
\end{picture}
}

\vbox{
\vspace{15mm}
\begin{equation}
  \label{eq:195c}
\hskip 30mm 
 - i \displaystyle
  \frac{g}{\sqrt{2}}\frac{m_{\ell}}{m_W}P_{R,L}
\end{equation}
\begin{picture}(0,5)
\psfrag{S}{$\varphi^\pm$}
\psfrag{f2}{$\nu, \ell$}
\psfrag{f1}{$\ell, \nu$}
\put(20,-3){\includegraphics[width=30mm]{PsiPsiHiggs.eps}}
\end{picture}
}

\subsection{Triple Higgs-Gauge and Goldstone-Gauge Interactions}

\vbox{
\vspace{13mm}
\begin{equation}
\hskip 20mm 
 -i\,\eta_e\, e\, \left( p_{+}-p_{-} \right)_\mu
\end{equation}
\begin{picture}(0,5)
\psfrag{A}{$A_\mu$}
\psfrag{s2}{$\varphi^+$}
\psfrag{s1}{$\varphi^-$}
\psfrag{p1}{$p_{-}$}
\psfrag{p2}{$p_{+}$}
\put(20,-5){\includegraphics[width=30mm]{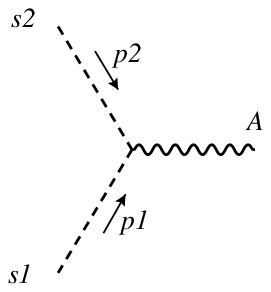}  }
\end{picture}
}

\vbox{
\vspace{15mm}
\begin{equation}
\hskip 30mm 
 -i\,\eta\,\eta_Z\, g\, \frac{\cos2\theta_W}{2\cos\theta_W}\left( p_{+}-p_{-} \right)_\mu
\end{equation}
\begin{picture}(0,5)
\psfrag{A}{$Z_\mu$}
\psfrag{s2}{$\varphi^+$}
\psfrag{s1}{$\varphi^-$}
\psfrag{p1}{$p_{-}$}
\psfrag{p2}{$p_{+}$}
\put(20,-3.5){\includegraphics[width=30mm]{PhiPhiGauge.eps}}
\end{picture}
}

\vbox{
\vspace{15mm}
\begin{equation}
\hskip 20mm 
 \pm \frac{i}{2}\,\eta\, g\,  \left( k-p \right)_\mu
\end{equation}
\begin{picture}(0,5)
\psfrag{A}{$W_\mu^\pm$}
\psfrag{s2}{$h$}
\psfrag{s1}{$\varphi^\mp$}
\psfrag{p1}{$\, k$}
\psfrag{p2}{$p$}
\put(20,-4){\includegraphics[width=30mm]{PhiPhiGauge.eps}}
\end{picture}
}

\vbox{
\vspace{15mm}
\begin{equation}
\hskip 20mm 
 -\eta\, \frac{g}{2}\,  \left( k-p \right)_\mu
\end{equation}
\begin{picture}(0,5)
\psfrag{A}{$W_\mu^\pm$}
\psfrag{s2}{$\!\!  \varphi_Z$}
\psfrag{s1}{$\varphi^\mp$}
\psfrag{p1}{$\, k$}
\psfrag{p2}{$p$}
\put(20,-4){\includegraphics[width=30mm]{PhiPhiGauge.eps}}
\end{picture}
}

\vbox{
\vspace{15mm}
\begin{equation}
\hskip 20mm 
-\eta\,\eta_Z\,  \frac{g}{2\cos\theta_W}\,  \left( k-p \right)_\mu
\end{equation}
\begin{picture}(0,5)
\psfrag{A}{$Z_\mu$}
\psfrag{s2}{$h$}
\psfrag{s1}{$\! \varphi_Z$}
\psfrag{p1}{$\, k$}
\psfrag{p2}{$p$}
\put(20,-4){\includegraphics[width=30mm]{PhiPhiGauge.eps}}
\end{picture}
}

\vbox{
\vspace{15mm}
\begin{equation}
\hskip 20mm 
  i\,\eta_e\eta\, e\, m_W\, g_{\mu\nu}
\end{equation}
\begin{picture}(0,6)
\psfrag{B}{$A_\mu$}
\psfrag{A}{$\!\! W_\nu^\pm$}
\psfrag{s}{$\varphi^\mp$}
\put(20,-6){\includegraphics[width=30mm]{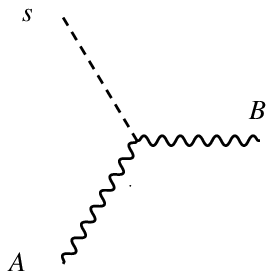}}
\end{picture}
}

\vbox{
\vspace{15mm}
\begin{equation}
\hskip 20mm 
  -i\,\eta_Z\, g\, m_Z\, \sin^2\theta_W\, g_{\mu\nu}
\end{equation}
\begin{picture}(0,6)
\psfrag{B}{$Z_\mu$}
\psfrag{A}{$\!\! W_\nu^\pm$}
\psfrag{s}{$\varphi^\mp$}
\put(20,-6){\includegraphics[width=30mm]{PhiGaugeGauge.eps}}
\end{picture}
}

\vbox{
\vspace{15mm}
\begin{equation}
\hskip 20mm 
  i g\, m_W\, g_{\mu\nu}
\end{equation}
\begin{picture}(0,6)
\psfrag{B}{$W_\mu^\pm$}
\psfrag{A}{$\!\! W_\nu^\mp$}
\psfrag{s}{$h$}
\put(20,-6){\includegraphics[width=30mm]{PhiGaugeGauge.eps}}
\end{picture}
}

\vbox{
\vspace{17mm}
\begin{equation}
\hskip 20mm 
 i\, \frac{ g}{\cos\theta_W}\, m_Z\, g_{\mu\nu}
\end{equation}
\begin{picture}(0,5)
\psfrag{B}{$Z_\mu$}
\psfrag{A}{$\!\! Z_\nu$}
\psfrag{s}{$h$}
\put(20,-4){\includegraphics[width=30mm]{PhiGaugeGauge.eps}}
\end{picture}
}

\subsection{Quartic Higgs-Gauge and Goldstone-Gauge Interactions}

\vbox{
\vspace{15mm}
\begin{equation}
  \label{eq:196}
\hskip 30mm
  \frac{i}{2}\, g^2\,  g_{\mu\nu}
\end{equation}
  \begin{picture}(0,5)
    \psfrag{H1}{$h$}
    \psfrag{H2}{$h$}
    \psfrag{V3}{$W_{\mu}^\pm$}
    \psfrag{V4}{$W_{\nu}^\mp$}
   \put(15,-4){\includegraphics[width=40mm]{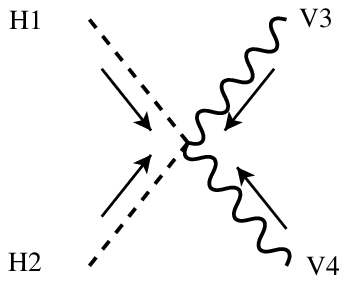}}
  \end{picture}
}

\vbox{
\vspace{15mm}
\begin{equation}
\hskip 30mm
  \frac{i}{2}\, g^2\,  g_{\mu\nu}
\end{equation}
  \begin{picture}(0,5)
    \psfrag{H1}{$\varphi_Z$}
    \psfrag{H2}{$\varphi_Z$}
    \psfrag{V3}{$W_{\mu}^\pm$}
    \psfrag{V4}{$W_{\nu}^\mp$}
    \put(15,-4){\includegraphics[width=40mm]{H1H2V3V4.eps}}
  \end{picture}
}

\vbox{
\vspace{15mm}
\begin{equation}
\hskip 30mm
  \frac{i}{2} \frac{g^2}{\cos^2\theta_W} g_{\mu\nu}
\end{equation}
  \begin{picture}(0,5)
    \psfrag{H1}{$h$}
    \psfrag{H2}{$h$}
    \psfrag{V3}{$Z_{\mu}$}
    \psfrag{V4}{$Z_{\nu}$}
    \put(15,-4){\includegraphics[width=40mm]{H1H2V3V4.eps}}
  \end{picture}
}

\vbox{
\vspace{15mm}
\begin{equation}
\hskip 30mm
  \frac{i}{2} \frac{g^2}{\cos^2\theta_W} g_{\mu\nu}
\end{equation}
  \begin{picture}(0,5)
    \psfrag{H1}{$\varphi_Z$}
    \psfrag{H2}{$\varphi_Z$}
    \psfrag{V3}{$Z_{\mu}$}
    \psfrag{V4}{$Z_{\nu}$}
    \put(15,-4){\includegraphics[width=40mm]{H1H2V3V4.eps}}
  \end{picture}
}

\vbox{
\vspace{15mm}
\begin{equation}
\hskip 30mm
  2 i\, e^2\,  g_{\mu\nu}
\end{equation}
  \begin{picture}(0,5)
    \psfrag{H1}{$\varphi^+$}
    \psfrag{H2}{$\varphi^-$}
    \psfrag{V3}{$A_{\mu}$}
    \psfrag{V4}{$A_{\nu}$}
    \put(15,-4){\includegraphics[width=40mm]{H1H2V3V4.eps}}
  \end{picture}
}

\vbox{
\vspace{15mm}
\begin{equation}
\hskip 30mm
  \frac{i}{2}\, \left(\frac{g \cos2\theta_W}{\cos\theta_W}\right)^2  g_{\mu\nu}
\end{equation}
  \begin{picture}(0,5)
    \psfrag{H1}{$\varphi^+$}
    \psfrag{H2}{$\varphi^-$}
    \psfrag{V3}{$Z_{\mu}$}
    \psfrag{V4}{$Z_{\nu}$}
    \put(15,-4){\includegraphics[width=40mm]{H1H2V3V4.eps}}
  \end{picture}
}

\vbox{
\vspace{15mm}
\begin{equation}
\hskip 30mm
  \frac{i}{2}\, g^2\,  g_{\mu\nu}
\end{equation}
  \begin{picture}(0,5)
    \psfrag{H1}{$\varphi^+$}
    \psfrag{H2}{$\varphi^-$}
    \psfrag{V3}{$W_{\mu}^+$}
    \psfrag{V4}{$W_{\nu}^-$}
    \put(15,-4){\includegraphics[width=40mm]{H1H2V3V4.eps}}
  \end{picture}
}

\vbox{
\vspace{15mm}
\begin{equation}
\hskip 30mm
  -i\,\eta_Z\, g^2\, \frac{\sin^2\theta_W}{2\cos\theta_W}\, g_{\mu\nu}
\end{equation}
  \begin{picture}(0,5)
    \psfrag{H1}{$\varphi^\mp$}
    \psfrag{H2}{$h$}
    \psfrag{V3}{$W_{\mu}^\pm$}
    \psfrag{V4}{$Z_{\nu}$}
    \put(15,-4){\includegraphics[width=40mm]{H1H2V3V4.eps}}
  \end{picture}
}

\vbox{
\vspace{15mm}
\begin{equation}
\hskip 30mm
  \mp\,\eta_Z\, g^2\, \frac{\sin^2\theta_W}{2\cos\theta_W}\, g_{\mu\nu}
\end{equation}
  \begin{picture}(0,5)
    \psfrag{H1}{$\varphi^\pm$}
    \psfrag{H2}{$\varphi_Z$}
    \psfrag{V3}{$W_{\mu}^\mp$}
    \psfrag{V4}{$Z_{\nu}$}
    \put(15,-4){\includegraphics[width=40mm]{H1H2V3V4.eps}}
  \end{picture}
}

\vbox{
\vspace{15mm}
\begin{equation}
\hskip 30mm
  \frac{i}{2}\, \eta_e \eta\, e g\, g_{\mu\nu}
\end{equation}
  \begin{picture}(0,5)
    \psfrag{H1}{$\varphi^\pm$}
    \psfrag{H2}{$h$}
    \psfrag{V3}{$W_{\mu}^\mp$}
    \psfrag{V4}{$A_{\nu}$}
    \put(15,-4){\includegraphics[width=40mm]{H1H2V3V4.eps}}
  \end{picture}
}

\vbox{
\vspace{15mm}
\begin{equation}
\hskip 30mm
  \mp \frac{1}{2}\, \eta_e \eta\, e g\, g_{\mu\nu}
\end{equation}
  \begin{picture}(0,5)
    \psfrag{H1}{$\varphi^\mp$}
    \psfrag{H2}{$\varphi_Z$}
    \psfrag{V3}{$W_{\mu}^\pm$}
    \psfrag{V4}{$A_{\nu}$}
    \put(15,-4){\includegraphics[width=40mm]{H1H2V3V4.eps}}
  \end{picture}
}

\vbox{
\vspace{15mm}
\begin{equation}
\hskip 30mm
   i\,\eta_e\eta\,\eta_Z\,  e g\, \frac{\cos2\theta_W}{\cos\theta_W}g_{\mu\nu}
\end{equation}
  \begin{picture}(0,5)
    \psfrag{H1}{$\varphi^+$}
    \psfrag{H2}{$\varphi^-$}
    \psfrag{V3}{$Z_{\mu}$}
    \psfrag{V4}{$A_{\nu}$}
    \put(15,-4){\includegraphics[width=40mm]{H1H2V3V4.eps}}
  \end{picture}
}

\subsection{Triple Higgs and Goldstone Interactions}

\vbox{
\vspace{15mm}
\begin{equation}
\hskip 30mm
   - \frac{i}{2}\, g\, \frac{m_h^2}{m_W}
\end{equation}
  \begin{picture}(0,5)
    \psfrag{s1}{$\varphi^-$}
    \psfrag{s2}{$\varphi^+$}
    \psfrag{s3}{$h$}
    \put(15,-7){\includegraphics[width=35mm]{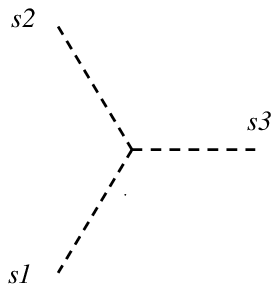}}
  \end{picture}
}

\vbox{
\vspace{20mm}
\begin{equation}
\hskip 30mm
-\frac{3}{2}\, i\, g\frac{m_h^2}{m_W} 
\end{equation}
  \begin{picture}(0,5)
    \psfrag{s1}{$h$}
    \psfrag{s2}{$h$}
    \psfrag{s3}{$h$}
    \put(15,-7){\includegraphics[width=35mm]{PhiPhiPhi.eps}}
  \end{picture}
}

\vbox{
\vspace{20mm}
\begin{equation}
\hskip 30mm
-\frac{i}{2}\,  g \frac{m_h^2}{m_W} 
\end{equation}
  \begin{picture}(0,5)
    \psfrag{s1}{$\varphi_Z$}
    \psfrag{s2}{$\varphi_Z$}
    \psfrag{s3}{$h$}
    \put(15,-7){\includegraphics[width=35mm]{PhiPhiPhi.eps}}
  \end{picture}
}

\subsection{Quartic Higgs and Goldstone Interactions}

\vbox{
\vspace{15mm}
\begin{equation}
\hskip 30mm
  -\frac{i}{2}\, g^2\, \frac{m_h^2}{m_W^2} 
\end{equation}
  \begin{picture}(0,5)
    \psfrag{H1}{$\varphi^+$}
    \psfrag{H2}{$\varphi^+$}
    \psfrag{H3}{$\varphi^-$}
    \psfrag{H4}{$\varphi^-$}
    \put(15,-4){\includegraphics[width=40mm]{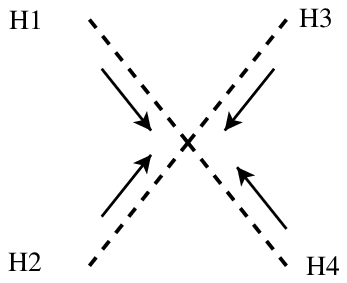}}
  \end{picture}
}

\vbox{

\vspace{15mm}
\begin{equation}
\hskip 30mm
  -\frac{i}{4}\, g^2\, \frac{m_h^2}{m_W^2} 
\end{equation}
  \begin{picture}(0,5)
    \psfrag{H1}{$\varphi^+$}
    \psfrag{H2}{$\varphi^-$}
    \psfrag{H3}{$h$}
    \psfrag{H4}{$h$}
    \put(15,-4){\includegraphics[width=40mm]{H1H2H3H4.eps}}
  \end{picture}
}

\vbox{
\vspace{15mm}
\begin{equation}
\hskip 30mm
  -\frac{i}{4}\, g^2\, \frac{m_h^2}{m_W^2} 
\end{equation}
  \begin{picture}(0,5)
    \psfrag{H1}{$\varphi^+$}
    \psfrag{H2}{$\varphi^-$}
    \psfrag{H3}{$\varphi_Z$}
    \psfrag{H4}{$\varphi_Z$}
    \put(15,-4){\includegraphics[width=40mm]{H1H2H3H4.eps}}
  \end{picture}
}

\vbox{
\vspace{15mm}
\begin{equation}
\hskip 30mm
  -\frac{3}{4}\, i\, g^2\frac{m_h^2}{m_W^2} 
\end{equation}
  \begin{picture}(0,5)
    \psfrag{H1}{$h$}
    \psfrag{H2}{$h$}
    \psfrag{H3}{$h$}
    \psfrag{H4}{$h$}
    \put(15,-4){\includegraphics[width=40mm]{H1H2H3H4.eps}}
  \end{picture}
}

\vbox{
\vspace{15mm}
\begin{equation}
\hskip 30mm
  -\frac{i}{4}\,  g^2\frac{m_h^2}{m_W^2} 
\end{equation}
  \begin{picture}(0,5)
    \psfrag{H1}{$\varphi_Z$}
    \psfrag{H2}{$\varphi_Z$}
    \psfrag{H3}{$h$}
    \psfrag{H4}{$h$}
    \put(15,-4){\includegraphics[width=40mm]{H1H2H3H4.eps}}
  \end{picture}
}

\vbox{
\vspace{15mm}
\begin{equation}
\hskip 30mm
  -\frac{3}{4}\, i\, g^2\frac{m_h^2}{m_W^2} 
\end{equation}
  \begin{picture}(0,5)
    \psfrag{H1}{$\varphi_Z$}
    \psfrag{H2}{$\varphi_Z$}
    \psfrag{H3}{$\varphi_Z$}
    \psfrag{H4}{$\varphi_Z$}
    \put(15,-4){\includegraphics[width=40mm]{H1H2H3H4.eps}}
  \end{picture}
}

\subsection{Ghost Propagators}

\vbox{
\vspace{3mm}
\begin{equation}
\hskip 30mm  \frac{\eta_G\, i}{k^2 + i \epsilon} 
\end{equation}
\begin{picture}(0,3)
\psfrag{G}{$c_A$}
\psfrag{k}{$k$}
\put(15,2){\includegraphics[width=3.5cm]{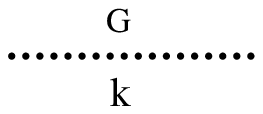}}
\end{picture}
}

\vbox{
\vspace{3mm}
\begin{equation}
\hskip 30mm  \frac{\eta_G\, i}{k^2 -\xi_W m_W^2 + i \epsilon} 
\end{equation}
\begin{picture}(0,3)
\psfrag{G}{$c^\pm$}
\psfrag{k}{$k$}
\put(15,2){\includegraphics[width=3.5cm]{GhostPropagatorFR-signs.eps}}
\end{picture}
}

\vbox{
\vspace{3mm}
\begin{equation}
\hskip 30mm  \frac{\eta_G\, i}{k^2 -\xi_Z m_Z^2 + i \epsilon} 
\end{equation}
\begin{picture}(0,3)
\psfrag{G}{$c_Z$}
\psfrag{k}{$k$}
\put(15,2){\includegraphics[width=3.5cm]{GhostPropagatorFR-signs.eps}}
\end{picture}
}

\subsection{Ghost Gauge Interactions}

\vbox{
\vspace{15mm}
\begin{equation}
\hskip 30mm \mp i\, \eta_G\, \eta_e\, e\, p_\mu
\end{equation}
\begin{picture}(0,5)
\psfrag{A}{$A_\mu$}
\psfrag{s1}{$c^\pm$}
\psfrag{s2}{$c^\pm$}
\psfrag{p}{$p$}
\put(15,-5){\includegraphics[width=30mm]{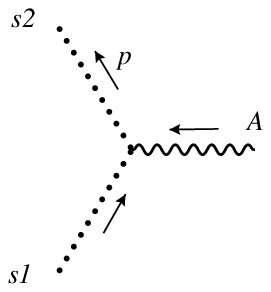}}  
\end{picture}
}

\vbox{
\vspace{15mm}
\begin{equation}
\hskip 30mm
\mp i\, \eta_G\, \eta\;\eta_Z\, g\,\cos\theta_W\, p_\mu
\end{equation}
\begin{picture}(0,5)
\psfrag{A}{$Z_\mu$}
\psfrag{s1}{$c^\pm$}
\psfrag{s2}{$c^\pm$}
\psfrag{p}{$p$}
\put(15,-6){\includegraphics[width=30mm]{GhostGhostGauge.eps}}  
\end{picture}
}

\vbox{
\vspace{15mm}
\begin{equation}
\hskip 30mm
\pm\, i\, \eta_G\, \eta\,\eta_Z\, g\,\cos\theta_W\, p_\mu
\end{equation}
\begin{picture}(0,5)
\psfrag{A}{$W_\mu^\pm$}
\psfrag{s2}{$c^\pm$}
\psfrag{s1}{$c_Z$}
\psfrag{p}{$p$}
\put(15,-6){\includegraphics[width=30mm]{GhostGhostGauge.eps}}  
\end{picture}
}

\vbox{
\vspace{15mm}
\begin{equation}
\hskip 30mm
\pm\, i\, \eta_G\, \eta_e\, e\, p_\mu
\end{equation}
\begin{picture}(0,5)
\psfrag{A}{$W_\mu^\pm$}
\psfrag{s2}{$c^\pm$}
\psfrag{s1}{$c_A$}
\psfrag{p}{$p$}
\put(15,-6){\includegraphics[width=30mm]{GhostGhostGauge.eps}}  
\end{picture}
}

\vbox{
\vspace{15mm}
\begin{equation}
\hskip 30mm
\pm\, i\, \eta_G\, \eta\, g\,\cos\theta_W\, p_\mu
\end{equation}
\begin{picture}(0,5)
\psfrag{A}{$W_\mu^\mp$}
\psfrag{s1}{$c^\pm$}
\psfrag{s2}{$c_Z$}
\psfrag{p}{$p$}
\put(15,-6){\includegraphics[width=30mm]{GhostGhostGauge.eps}}  
\end{picture}
}

\vbox{
\vspace{15mm}
\begin{equation}
\hskip 30mm
\pm\, i\, \eta_G\, \eta_e\, e\, p_\mu
\end{equation}
\begin{picture}(0,5)
\psfrag{A}{$W_\mu^\mp$}
\psfrag{s1}{$c^\pm$}
\psfrag{s2}{$c_A$}
\psfrag{p}{$p$}
\put(15,-6){\includegraphics[width=30mm]{GhostGhostGauge.eps}}  
\end{picture}
}

\subsection{Ghost Higgs and Ghost Goldstone Interactions}

\vbox{
\vspace{15mm}
\begin{equation}
\hskip 30mm
\pm\,  \eta_G\, \frac{g}{2}\, \xi_W\, m_W
\end{equation}
\begin{picture}(0,5)
\psfrag{A}{$\varphi_Z$}
\psfrag{s1}{$c^\pm$}
\psfrag{s2}{$c^\pm$}
\put(15,-5){\includegraphics[width=30mm]{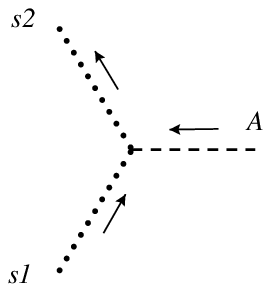}}  
\end{picture}
}

\vbox{
\vspace{15mm}
\begin{equation}
\hskip 30mm
-\frac{i}{2}\,  \eta_G\, g\, \xi_W\, m_W
\end{equation}
\begin{picture}(0,5)
\psfrag{A}{$h$}
\psfrag{s1}{$c^\pm$}
\psfrag{s2}{$c^\pm$}
\put(15,-5){\includegraphics[width=30mm]{GhostGhostHiggs.eps}}  
\end{picture}
}

\vbox{
\vspace{15mm}
\begin{equation}
\hskip 30mm
- \eta_G\, \frac{i g}{2\cos\theta_W}\, \xi_Z\,  m_Z
\end{equation}
\begin{picture}(0,5)
\psfrag{A}{$h$}
\psfrag{s1}{$c_Z$}
\psfrag{s2}{$c_Z$}
\put(15,-5){\includegraphics[width=30mm]{GhostGhostHiggs.eps}}  
\end{picture}
}

\vbox{
\vspace{15mm}
\begin{equation}
\hskip 30mm
\frac{i}{2}\, \eta_G\, \eta_Z\, g\, \xi_Z\,  m_Z
\end{equation}
\begin{picture}(0,5)
\psfrag{A}{$\varphi^\mp$}
\psfrag{s1}{$c^\pm$}
\psfrag{s2}{$c_Z$}
\put(15,-5){\includegraphics[width=30mm]{GhostGhostHiggs.eps}}  
\end{picture}
}

\vbox{
\vspace{15mm}
\begin{equation}
\hskip 30mm
-i\,  \eta_G\, \eta_Z\, g\, \frac{\cos2\theta_W}{2\cos\theta_W}\, \xi_W\,  m_W
\end{equation}
\begin{picture}(0,5)
\psfrag{A}{$\varphi^\pm$}
\psfrag{s2}{$c^\pm$}
\psfrag{s1}{$c_Z$}
\put(15,-5){\includegraphics[width=30mm]{GhostGhostHiggs.eps}}  
\end{picture}
}

\vbox{
\vspace{18mm}
\begin{equation}
\hskip 30mm
-i\, \eta_G\, \eta_e\,\eta\, e\,  \xi_W\, m_W
\end{equation}
\begin{picture}(0,5)
\psfrag{A}{$\varphi^\pm$}
\psfrag{s2}{$c^\pm$}
\psfrag{s1}{$c_A$}
\put(15,-5){\includegraphics[width=30mm]{GhostGhostHiggs.eps}}  
\end{picture}
}


\subsection{Brief comment on the alternative metric}

As mentioned in the introduction,
all our calculations and Feynman diagrams have
been obtained with the metric $(+,-,-,-)$.
A few books use instead the metric $(-,+,+,+)$.
Ref.~\refcite{Srednicki:2007qs} differs from ours
only in the metric.
For example,
it uses as we do,
$i \bar{\psi} \slash{\partial} \psi$ for the fermion kinetic term.
Therefore,
our results agree,
with the change $g_{\mu \nu} \rightarrow - g_{\mu \nu}$,
implying also changes of the type
$p^2 \rightarrow - p^2$ and $\slash{p} \rightarrow - \slash{p}$.
The comparison is much more involved with respect
to Refs.~\refcite{Weinberg:1996kr} and \refcite{Bardin:1999ak},
because in those cases there are many changes
besides the metric,
involving,
in particular,
multiple factors of $i$ and $2 \pi$ in the Feynman rules.
As an additional complication,
Ref.~\refcite{Bardin:1999ak} uses
$- \bar{\psi} \slash{\partial} \psi$ for the fermion kinetic term,
implying also a change in the matrices $\gamma^\mu$,
compounded by different gauge fixing terms.
A detailed analysis of all such choices lies
beyond the scope of this work.


\section*{Acknowledgments}

We are grateful to A.~Barroso and P.~Nogueira for useful
discussions and to H.~Ser\^{o}dio for reading the manuscript
and making suggestions.
This work was funded by FCT through the projects
CERN/FP/116328/2010, PTDC/FIS/098188/2008 and U777-Plurianual,
and by the EU RTN project Marie Curie: PITN-GA-2009-237920. JCR also
acknowledge support from project PTDC/FIS/102120/2008.
\appendix

\section{\label{app:BRST}Gauge and BRST consistency checks}

\subsection{Gauge transformation and gauge invariance}

For completeness we write here the gauge transformations of the gauge
fixing terms needed to find the Lagrangian in Eq.~(\ref{eq:107}).
It is convenient to redefine the parameters as
\begin{eqnarray}
\alpha^\pm &=&
\frac{\alpha^1 \mp \alpha^2}{\sqrt{2}} ,
\nonumber\\[+2mm]
\eta_Z \alpha_Z &=&
\alpha^3 \cos\theta_W -\eta_{\theta} \alpha^4 \sin\theta_W ,
\nonumber\\[+2mm]
\alpha_A &=&
\eta_{\theta} \alpha^3 \sin\theta_W + \alpha^4 \cos\theta_W .
\end{eqnarray}
We then get
\begin{eqnarray}
\delta F_G^a
&=&
\partial^\mu\left(
- \partial_\mu \beta^a - \eta_s g_s f^{abc} \beta^b G_\mu^c\right) ,
\nonumber\\[+2mm]
\delta F_A
&=&
\partial^\mu(\delta A^\mu) ,
\nonumber\\[+2mm]
\delta F_Z
&=&
\partial_\mu (\delta Z^\mu)  + \eta\,\eta_Z\,\xi_Z m_Z \delta \varphi_Z ,
\nonumber\\[+2mm]
\delta F_{+}
&=& \partial_\mu (\delta W_\mu^+ ) + i\,\eta\, \xi_W m_W \delta \varphi^+  ,
\nonumber\\[+2mm]
\delta F_{-}
&=&
\partial_\mu (\delta W_\mu^- ) - i\,\eta\, \xi_W m_W \delta \varphi^- .
\label{eq:110}
\end{eqnarray}
Using the explicit form of the gauge transformations we can finally
find the missing pieces,
\begin{eqnarray}
\delta A_\mu
&=&
-\partial_\mu \alpha_A -i\,\eta_e\, e\,
\left(W_\mu^+ \alpha^- - W_\mu^- \alpha^+ \right) ,
\nonumber\\[+2mm]
\delta Z_\mu
&=&
-\partial_\mu \alpha_Z - i\,\eta\,\eta_Z\, g \cos\theta_W 
\left( W_\mu^+ \alpha^- - W_\mu^- \alpha^+ \right)  ,
\nonumber\\[+2mm]
\delta W_\mu^+
&=&
-\partial_\mu \alpha^+ - i\,\eta\, g 
\left[ \alpha^+ \left(\eta_Z\, Z_\mu \cos\theta_W 
+ \eta_{\theta}A_\mu \sin\theta_W \right)
\right.
\nonumber\\[+2mm]
& &
\left.
  - \left( \eta_Z\, \alpha_Z \cos\theta_W 
+\eta_{\theta} \alpha_A \sin\theta_W \right)  W_\mu^+
  \right] ,
\nonumber\\[+2mm]
\delta W_\mu^-
&=&
-\partial_\mu \alpha^- + i\,\eta\,  g 
\left[ \alpha^- \left(\eta_Z\,Z_\mu \cos\theta_W 
+ \eta_{\theta}A_\mu \sin\theta_W \right)
\right.
\nonumber\\[+2mm]
& &
\left.
  - \left( \eta_Z\,\alpha_Z \cos\theta_W 
+ \eta_{\theta}\alpha_A \sin\theta_W \right)  W_\mu^-
  \right].
\label{eq:111}
\end{eqnarray}
To get the variation of the Goldstone bosons we notice that 
\begin{eqnarray}
\delta \Phi
&=&
\left[  i\, \eta \frac{ g}{\sqrt{2}} \left(
 \tau^+ \alpha^+ +  \tau^- \alpha^- \right) 
+i\, \eta \frac{g}{2} \tau_3 \alpha^3
+ i\, \eta' \frac{g'}{2} \alpha^4 \right] \Phi
\\[+2mm]
&=&
\left[  i\,\eta \frac{g}{\sqrt{2}} \left(
 \tau^+ \alpha^+ +  \tau^- \alpha^- \right) +i\, \eta_e e\, Q\,   \alpha_A
\right.
\nonumber\\
&&
\ \ 
\left.
+\ i\,\eta \frac{g}{\cos\theta_W} \left(\frac{\tau_3}{2} - Q\, \sin^2\theta_W
\right)\eta_Z \alpha_Z \right] \Phi
 \label{eq:1}
\end{eqnarray}
which we can write as
\begin{equation}
  \label{eq:2}
  \begin{bmatrix}
    \delta\varphi^+\\[+2mm]
\displaystyle
    \frac{\delta(H + i \varphi_Z)}{\sqrt{2}}\\ 
  \end{bmatrix}
=-\frac{i}{2}
\begin{bmatrix}
\displaystyle
  -\eta\,g \tfrac{\cos2\theta_W}{\cos\theta_W} \eta_Z \alpha_Z -2
  \eta_e e \alpha_A &
  -\sqrt{2} \eta\, g \alpha^+ \\[+2mm]
\displaystyle
 -\sqrt{2} \eta\, g \alpha^- & \eta \tfrac{g}{\cos\theta_W} \eta_Z
 \alpha_Z \\
\end{bmatrix}
\begin{bmatrix}
  \varphi^+\\[+2mm]
\displaystyle
 \frac{v + H + i \varphi_Z}{\sqrt{2}}
\end{bmatrix} ,
\end{equation}
leading to
\begin{eqnarray}
\delta\varphi_Z
&=&
\displaystyle
  \frac{1}{2}\eta\, g \left(\alpha^- \varphi^+ +
\alpha^+ \varphi^- \right) -\eta \frac{g}{2 \cos\theta_W}\eta_Z\,
 \alpha_Z (v + H) ,
\nonumber\\[+2mm]
\delta \varphi^+
&=&
\displaystyle
i\,\eta\, \frac{g}{2} (v + H + i \varphi_Z) \alpha^+ 
+ i\,\eta\, \frac{g}{2}  \frac{\cos 2\theta_W}{\cos\theta_W} \varphi^+
\eta_Z\, \alpha_Z
+ i\, \eta_e e\, \varphi^+ \alpha_A ,
\nonumber\\[+2mm]
\delta \varphi^-
&=&
\displaystyle
 -i\,\eta\, \frac{g}{2} (v + H - i \varphi_Z) \alpha^- 
- i\,\eta\, \frac{g}{2}  \frac{\cos 2\theta_W}{\cos\theta_W} \varphi^-
\eta_Z\, \alpha_Z - i\,\eta_e  e\, \varphi^- \alpha_A ,
\nonumber\\[+2mm]
\delta H
&=& 
-i\,\eta\, \frac{g}{2} ( \alpha^+ \varphi^- - \alpha^- \varphi^+)  
+ \eta\, \frac{g}{2\cos\theta_W} \eta_Z\, \alpha_Z\, \varphi_Z .
\label{eq:112}
\end{eqnarray}

With the gauge transformations given in Eqs.~(\ref{eq:111}) and
(\ref{eq:112}),
one can easily verify that $\mathcal{L}_{\rm gauge}$
and  $\mathcal{L}_{\rm Higgs}$ are gauge
invariant, independently of the choice of the $\eta$'s.
For instance,
for $\mathcal{L}_{\rm Higgs}$ we have 
\begin{equation}
  \label{eq:3}
  \delta \mathcal{L}_{\rm Higgs} = \delta \left(D_\mu
    \Phi\right)^\dagger D^\mu \Phi + \left(D_\mu
    \Phi\right)^\dagger \delta \left(D^\mu \Phi\right)
+ \delta \left( \mu^2 \Phi^\dagger \Phi - \lambda \left(\Phi^\dagger \Phi
\right)^2 \right)
 = 0 \ .
\end{equation}
To check the fermion part we have to give explicitly the gauge
transformations for $\psi_L$ and $\psi_R$.
They can be easily obtained
from Eqs.~(\ref{eq:115}) and (\ref{eq:105}).
We get,
\begin{eqnarray}
\delta \psi_L
&=&
\left[   i\, \eta \frac{g}{\sqrt{2}} \left(
 \tau^+ \alpha^+  +  \tau^- \alpha^- \right) +i\, \eta_e e\, Q\,   \alpha_A
\right.
\nonumber\\
&&
\left.
\hspace{5mm}
+ i\, \eta \frac{g}{\cos\theta_W} \left(\frac{\tau_3}{2} - Q\, \sin^2\theta_W
\right) \eta_Z \alpha_Z \right] \psi_L ,
\nonumber\\
\delta \psi_R
&=&
\left[ i\, \eta_e  e\, Q\,   \alpha_A
- i\, \eta \frac{ g}{\cos\theta_W}  Q\, \sin^2\theta_W 
\eta_Z \alpha_Z \right] \psi_R\, ,
\label{eq:5}
\end{eqnarray}
supplemented by Eq.~\eqref{eq:121} for the $SU(3)_c$
transformation of the quarks.
Using these transformation laws one can verify that
\begin{equation}
  \label{eq:6}
\delta \left( \mathcal{L}_{\rm Fermion} + \mathcal{L}_{\rm
    Yukawa}\right) =0\ ,
\end{equation}
completing the proof of the gauge invariance of the classical part of 
$\mathcal{L}_{\rm SM}$.
This means that,
except for $\mathcal{L}_{\rm GF} + \mathcal{L}_{\rm Ghost}$
to be discussed below,
we have included the various
$\eta$ parameters in the appropriate fashion.

\subsection{Consistency checks and the BRST transformations}

To make the proof of gauge invariance for the complete Lagrangian
we have to deal with
the gauge fixing and ghost terms. This is more easily done using the
BRST transformations of Becchi, Rouet and Stora\cite{becchi:1975nq} and
Tyutin\cite{tyutin:1975qk},
and the Slavnov operator.
The Slavnov operator is a special kind of gauge transformation
on the gauge and matter fields.
More specifically,
we define,
\begin{equation}
  \label{eq:7}
  s(A_\mu^i) = \frac{\delta A_\mu^i}{\delta \alpha_i} \ c^i, \quad
  s(\phi_i) = \frac{\delta \phi_i}{\delta \alpha_i} \ c^i\ ,
\end{equation}
where $A_\mu^i$ are the gauge fields and $\phi_i$ represents
generically any matter field (fermion or boson). They have the
following properties:

\begin{romanlist}[(ii)]
\item For a product of two fields, we have
  \begin{equation}
    \label{eq:9}
    s( X Y) = s(X) Y + (-1)^{{\rm GN}(X)} X s(Y)  \ .
  \end{equation}
In this expression the ghost number,
 ${\rm GN}(X)$, is defined as zero for gauge and matter fields, $+1$ for
$c^i$ fields (ghosts) and $-1$ for $\bar{c}^i$ (anti-ghosts).

\item $s$ raises the dimension by one unit (in terms of mass).

\item $s$ does not change the charge.

\item The Slavnov operator is nilpotent, that is, $s^2 =0$.

\end{romanlist}

To check the last identity we must have, for a non-abelian group,
\begin{equation}
  \label{eq:8}
  s(c^i) = -\eta\, \frac{g}{2} f^{ijk} c^j c^k\ .
\end{equation}
Let us show how the nilpotency of $s$ is obtained for the gauge fields
of a non-abelian theory. From Eq.~(\ref{eq:7}) we have,
\begin{equation}
  \label{eq:11}
  s(A_\mu^i) = -\partial_\mu c^i - \eta\, g f^{ijk} c^j A_\mu^k\ .
\end{equation}
Therefore, using Eq.~(\ref{eq:9}) we get,
\begin{eqnarray}
  s^2 A_\mu^i
&=&
-\partial_\mu s(c^i) - \eta\, g f^{ijk} s(c^j)A_\mu^k
  + \eta\, g f^{ijk} c^j s(A_\mu^k)
\nonumber\\
&=&
\eta\, \, \frac{g}{2}f^{ijk} \left( \partial_\mu c^j c^k +
    c^j \partial_\mu c^k \right)
+ \eta^2 \frac{g^2}{2} f^{ijk}
f^{jmn} c^m c^n A_\mu^k
\nonumber\\
&&
+\, \eta\, g f^{ijk} c^j \left(-\partial_\mu c^k
- \eta\, g\, f^{kmn} c^m A_\mu^n\right)
\nonumber\\
&=&
\eta\, g\, f^{ijk} \left( c^j \partial_\mu c^k - c^j \partial_\mu
  c^k \right)
+ \frac{g^2}{2} \left(f^{ijk} f^{jmn} +  f^{ijm} f^{jnk}
+  f^{ijn} f^{jkm} \right)c^m c^n
A_\mu^k
\nonumber\\
&=&
0\ ,
\end{eqnarray}
where we have used the anti-symmetry of the structure constants and of
the ghost fields, and the Jacobi identity. This confirms that the
assignment of Eq.~(\ref{eq:8}) is consistent with Eqs.~(\ref{eq:9})
and (\ref{eq:11}).
Before proceeding,
we should notice that another definition for the
product can be used.
In particular,
Ref.~\refcite{romao:1986vp} uses
\begin{equation}
  \label{eq:10}
  s( X Y) = (-1)^{{\rm GN}(Y)} s(X) Y +  X s(Y)  \ .
\end{equation}
Then, to verify the nilpotency of $s$, we must reverse the sign in
Eq.~(\ref{eq:8}). 

To prove the invariance of $\mathcal{L}_{\rm GF + Ghost}$ we use the
BRST technique. This is best explained for a simple group. We have,
\begin{equation}
  \label{eq:12}
  \mathcal{L}_{\rm GF + Ghost} = -\frac{1}{2\xi} F_i^2
+ \eta_G\, \overline{c}^j\,
  \frac{\delta F_j}{\delta \alpha_i} c^i =
-\frac{1}{\xi} F_i^2 + \overline{c}^j\,
  s(F_j) \ ,
\end{equation}
where the last step follows from Eq.~(\ref{eq:7}). Now, because of the
nilpotency of the Slavnov operator, to ensure the
invariance of Eq.~(\ref{eq:12}) under BRST transformations it is
enough to require that
\begin{equation}
  \label{eq:13}
  s(\overline{c}^j) = \eta_G\, \frac{1}{\xi} F^j\ .
\end{equation}
If the gauge fixing is non-linear,
some subtleties arise,
as explained in Ref.~\refcite{romao:1986vp}.

Coming back to the Standard Model,
we only have to verify that the
Slavnov operator is indeed nilpotent in all the fields.
We have verified this explicitly for all the cases.
For completeness,
we give
here the action of the Slavnov operator in all of the Standard Model
fields,
in a way consistent with our notation.
We just give the electroweak part, because, for QCD,
they can be read from Eqs.~(\ref{eq:8}) and (\ref{eq:11}). 
We start with the gauge fields,
\begin{eqnarray}
s(A_\mu)
&=&
-\partial_\mu c_A -i\,\eta_e\, e\,
 \left(W_\mu^+ c^- - W_\mu^- c^+ \right),
\nonumber \\[+2mm]
s(Z_\mu)
&=&
-\partial_\mu c_Z - i\,\eta\,\eta_Z\, g \cos\theta_W 
\left( W_\mu^+ c^- - W_\mu^- c^+ \right),
\nonumber\\[+2mm]
s(W_\mu^+)
&=&
-\partial_\mu c^+ - i\,\eta\, g 
\left[ c^+ \left(\eta_Z\, Z_\mu \cos\theta_W 
+ \eta_{\theta}A_\mu \sin\theta_W \right)
\right.
\nonumber\\[+2mm]
&&
\left.
- \left( \eta_Z\, c_Z \cos\theta_W 
+\eta_{\theta} c_A \sin\theta_W \right)  W_\mu^+
\right],
\nonumber\\[+2mm]
s(W_\mu^-)
&=&
-\partial_\mu c^- + i\,\eta\,  g 
\left[ c^- \left(\eta_Z\,Z_\mu \cos\theta_W 
+ \eta_{\theta}A_\mu \sin\theta_W \right) 
\right.
\nonumber\\[+2mm]
&&
\left.
- \left( \eta_Z\,c_Z \cos\theta_W 
+ \eta_{\theta}c_A \sin\theta_W \right)  W_\mu^-
  \right] \ .
\label{eq:14}
\end{eqnarray}
For the Higgs we get
\begin{eqnarray}
s(\varphi_Z)
&=&
\displaystyle
  \frac{1}{2}\eta\, g \left(c^- \varphi^+ +
c^+ \varphi^- \right) -\eta \frac{g}{2 \cos\theta_W}\eta_Z\,
 c_Z (v + H),
\nonumber\\[+2mm]
s(\varphi^+)
&=&
\displaystyle
i\,\eta\, \frac{g}{2} (v + H + i \varphi_Z) c^+ 
+ i\,\eta\, \frac{g}{2}  \frac{\cos 2\theta_W}{\cos\theta_W} \varphi^+
\eta_Z\, c_Z
+ i\, \eta_e e\, \varphi^+ c_A,
\nonumber\\[+2mm]
s(\varphi^-)
&=&
\displaystyle
 -i\,\eta\, \frac{g}{2} (v + H - i \varphi_Z) c^- 
- i\,\eta\, \frac{g}{2}  \frac{\cos 2\theta_W}{\cos\theta_W} \varphi^-
\eta_Z\, c_Z - i\,\eta_e  e\, \varphi^- c_A,
\nonumber\\[+2mm]
s(H)
&=& 
-i\,\eta\, \frac{g}{2} ( c^+ \varphi^- - c^- \varphi^+)  
+ \eta\, \frac{g}{2\cos\theta_W} \eta_Z\, c_Z\, \varphi_Z\ ,
\label{eq:15}
\end{eqnarray}
and for the fermions
\begin{eqnarray}
s(\psi_L) 
&=&
\left[   i\, \eta \frac{g}{\sqrt{2}} \left(
 \tau^+ c^+  +  \tau^- c^- \right) +i\, \eta_e e\, Q\,   c_A
\right.
\nonumber\\
&&
\left.
\hspace{4mm}
+ i\, \eta \frac{g}{\cos\theta_W} \left(\frac{\tau_3}{2} - Q\, \sin^2\theta_W
\right) \eta_Z c_Z \right] \psi_L,
\nonumber\\
s(\psi_R)
&=& \left[ i\, \eta_e  e\, Q\,   c_A
- i\, \eta \frac{ g}{\cos\theta_W}  Q\, \sin^2\theta_W 
\eta_Z c_Z \right] \psi_R\, .
\label{eq:16}
\end{eqnarray}
Finally,
we need the rules for the ghost fields.
These are obtained from Eq.~(\ref{eq:8}).
We get,
\begin{eqnarray}
s(c_A)
&=&
i\, \eta_e\, e\, c^+ c^- ,
\nonumber\\[+2mm]
s(c_Z)
&=&
i\, \eta\, \eta_Z g \cos\theta_W\ c^+ c^- ,
\nonumber\\[+2mm]
s(c^+)
&=&
i\, \eta\,\eta_Z\, g \cos\theta_W\ c_Z c^+ 
         + i\, \eta_e\, e\, c_A c^+ ,
\nonumber\\[+2mm]
s(c^-)
&=&
-i\,\eta\,\eta_Z\, g  \cos\theta_W\ c_Z c^- 
- i\, \eta_e e\, c_A c^-\ .
\label{eq:17}
\end{eqnarray}

%


\end{document}